\documentclass[a4paper,11pt,preprint]{article}
\pdfoutput=1 

\usepackage{jheppub} 

\usepackage[T1]{fontenc} 

\usepackage{graphicx}
\usepackage{amsmath}
\usepackage{amssymb}

\def\ba{\begin{eqnarray}}
\def\ea{\end{eqnarray}}

\def\be{\begin{equation}}
\def\ee{\end{equation}}

\newcommand{\checked}[1]{}

\newcommand{\la}{\label}
\def\sumint{\hbox{$\sum$}\!\!\!\!\!\!\!\int}

\title{Two-loop HTL-resummed thermodynamics for $\mathcal{N}=4$ supersymmetric Yang-Mills theory}

\author[a,b]{Qianqian Du,}
\author[b]{Michael Strickland,}
\author[b]{Ubaid Tantary,}
\author[a]{and Ben-Wei Zhang}
\affiliation[a]{Institute of Particle Physics and Key Laboratory of Quark and Lepton Physics (MOS), Central China Normal University, Wuhan, 430079, China}
\affiliation[b]{Department of Physics, Kent State University, Kent, OH 44242, United States}
\emailAdd{duqianqianstudent@mails.ccnu.edu.cn}
\emailAdd{mstrick6@kent.edu}
\emailAdd{utantary@kent.edu}
\emailAdd{bwzhang@mail.ccnu.edu.cn}

\abstract{
We compute the two-loop hard-thermal-loop (HTL) resummed thermodynamic potential for $\mathcal{N}=4$ supersymmetric Yang-Mills (SYM). Our final result is manifestly gauge-invariant and was renormalized using only simple vacuum energy, gluon mass, scalar mass, and quark mass counter terms.  The HTL mass parameters $m_D$, $M_D$, and $m_q$ are then determined self-consistently using a variational prescription which results in a set of coupled gap equations.  Based on this, we obtain the two-loop HTL-resummed thermodynamic functions of $\mathcal{N}=4$ SYM.  We compare our final result with known results obtained in the weak- and strong-coupling limits.  We also compare to previously obtained approximately self-consistent HTL resummations and Pad\'{e} approximants.  We find that the two-loop HTL resummed results for the scaled entropy density is a quantitatively reliable approximation to the scaled entropy density for $0 \leq \lambda \lesssim 2$ and is in agreement with previous approximately self-consistent HTL resummation results for $\lambda \lesssim 6$.
}

\keywords{high-temperature perturbation theory, supersymmetric Yang-Mills, diagrammatic resummation, hard thermal loops}

\begin{document}

\setcounter{tocdepth}{1}

\maketitle
\flushbottom

\section{Introduction}
\la{sect:intro}

$\mathcal{N}=4$ supersymmetric Yang-Mills theory (SYM) is the most famous example of a conformal field theory (CFT) in four dimensions, and is often taken as a model for hot QCD in the large number of colors $N_c$ and strong 't Hooft coupling $\lambda = g^2 N_c$ limits. The strong coupling behavior of the free energy has been computed using the anti-de Sitter space/CFT (AdS/CFT) correspondence \cite{Gubser:1998nz}, with the result being
\ba\la{stro}
\frac{\mathcal{F}}{\mathcal{F}_{\textrm{ideal}}}= \frac{\mathcal{S}}{\mathcal{S}_{\textrm{ideal}}} =\frac{3}{4}\bigg[1+\frac{15}{8}\zeta(3)\lambda^{-3/2} + \mathcal{O}(\lambda^{-2}) \bigg] ,
\ea
where $\mathcal{F}_{\textrm{ideal}} = - d_A \pi^2T^4/6$ is the ideal or Stefan-Boltzmann limit of the free energy and $\mathcal{S}_{\textrm{ideal}} = 2 d_A \pi^2T^3/3$, with $d_A = N_c^2 -1$ being the dimension of the adjoint representation.

In the weak-couping limit the $\mathcal{N}=4$ SYM free energy has been calculated through order  $\lambda^{3/2}$ giving~\cite{Fotopoulos:1998es,Kim:1999sg,VazquezMozo:1999ic}
\ba\la{weaexp}
\frac{\mathcal{F}}{\mathcal{F}_{\textrm{ideal}}}= \frac{\mathcal{S}}{\mathcal{S}_{\textrm{ideal}}} =1-\frac{3}{2\pi^2}\lambda+\frac{3+\sqrt{2}}{\pi^3}\lambda^{3/2} + \mathcal{O}(\lambda^2)   \, .
\ea
Note that, since the beta function of the $\mathcal{N}=4$ SYM theory is zero, the coupling constant does not run and is independent of the temperature.  As a result, we can vary the coupling between the two limits at each temperature.

\begin{figure}[t!]
\centering
\includegraphics[width=0.9\textwidth]{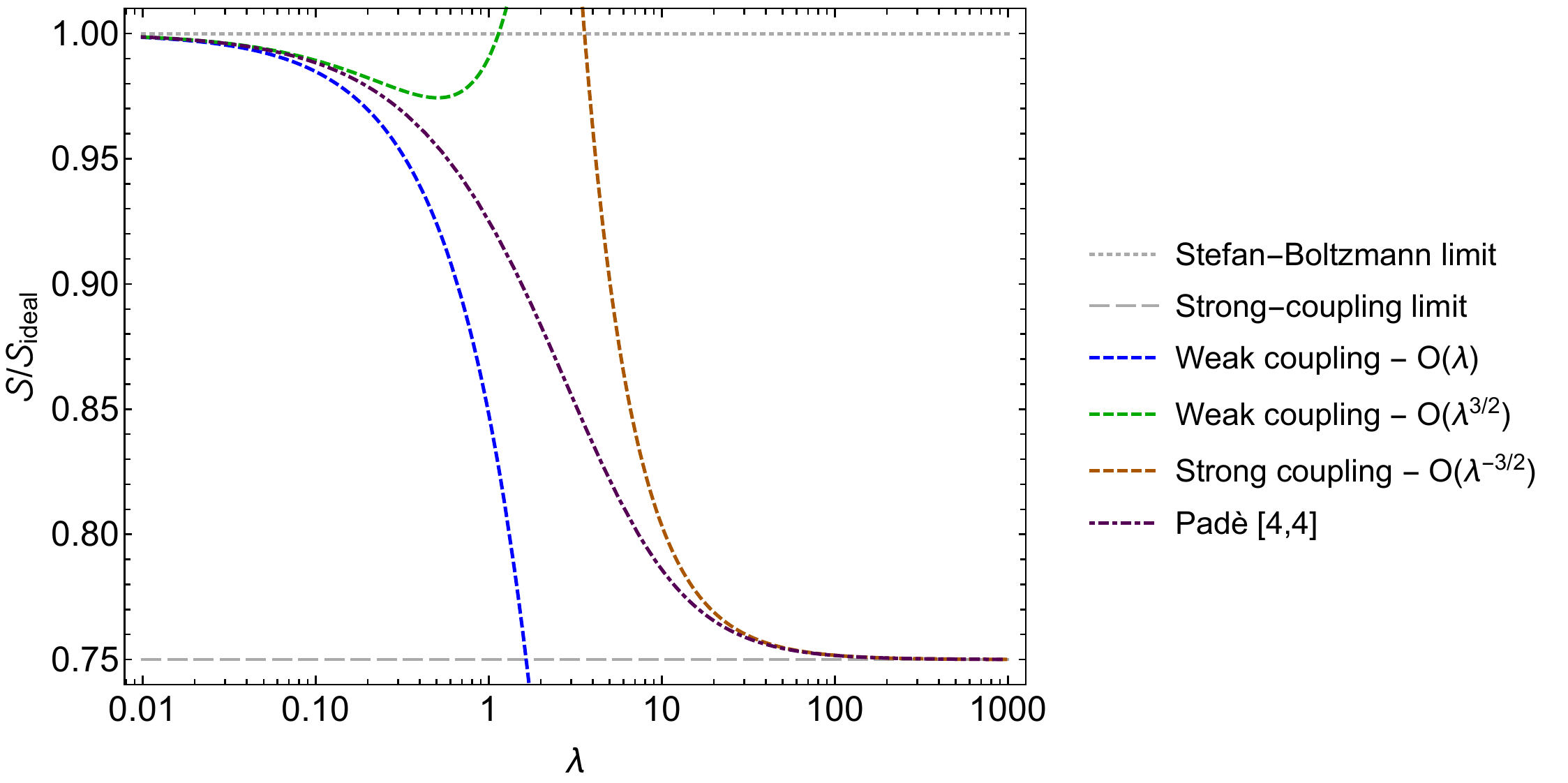}
\caption{\label{introFigure}
Weak and strong coupling results for the entropy density in $\mathcal{N}=4$ SYM theory compared to a $R_{[4,4]}$ Pad\`{e} approximation constructed from both limits.}
\end{figure}

One expects these two series to describe their respective asymptotic limits correctly, however, the radius of convergence of each of these series is unknown and, therefore, it is unclear to what degree each of these can trusted away from their respective limits.   In Fig.~\ref{introFigure} we plot the scaled entropy density resulting from Eqs.~\eqref{stro} and \eqref{weaexp} as a function of $\lambda$ along with a $R_{[4,4]}$ Pad\'{e} approximant constructed from these results \cite{Kim:1999sg,Blaizot:2006tk}.\footnote{Although a Pad\'{e} approximant might provide a convenient interpolation between the weak- and strong-coupling limits their construction is in no sense systematic.  In particular, the resulting expressions are incomplete since we know that, at least at weak coupling, the series will contain logarithms of the coupling constant beyond ${\mathcal O}(\lambda^{3/2})$.}  From this Figure, we can see that the two successive weak coupling approximations are only close to one another below $\lambda \sim 0.1$ and rapidly diverge beyond $\lambda \sim 1$.  In the strong coupling limit, only the first two terms in the series are known.  As can be seen from Fig.~\ref{introFigure} the strong coupling result diverges quickly below $\lambda \sim  10$.  The question then becomes, how can we systematically extend these two results into the {\em intermediate coupling} region $\lambda \sim 1 - 10$.  In this paper, we present progress towards this goal in the weak-coupling limit using hard-thermal-loop (HTL) perturbation theory.  Our study is complementary to the earlier work of Blaizot, Iancu, and Rebhan in which they applied HTL resummation using an approximately self-consistent scheme \cite{Blaizot:2006tk}.  The key difference from this earlier work is that our result is manifestly gauge invariant and based on the systematically improvable HTL perturbation theory (HTLpt) framework \cite{Andersen:1999fw,Andersen:1999sf,Andersen:1999va}.

The goal of our work is to improve the convergence of the successive weak-coupling approximations. One promising approach is to use a variational framework in which the free energy $\mathcal{F}$ is expressed as the variational minimum of the thermodynamic potential $\Omega(T, \lambda; m^2)$ that depends on one or more variational parameters that we denote collectively by ${\bf a}$
\ba\la{fmin}
\mathcal{F}(T,\lambda)=\Omega(T, \lambda; {\bf a})\bigl|_{\partial \Omega/\partial {\bf a}=0} \,.
\ea
For example, the $\Phi$-derivable approximation is a widely used variational method in which the propagator is used as an infinite set of variational parameters~\cite{Luttinger:1960ua,Baym:1962sx}. The $\Phi$-derivable thermodynamic potential $\Omega$ is given by the $2$-particle-irreducible (2PI) effective action, which is the sum of all diagrams that are $2$-particle-irreducible with respect to the complete propagator~\cite{Cornwall:1974vz}.  This method is difficult to apply to relativistic field theories except for the case where the self-energy is momentum-independent.  Despite this, there still have some progress in applications to quantum chromodynamics (QCD) and electrodynamics (QED)~\cite{Braaten:2001en,Braaten:2001vr,vanHees:2001ik,vanHees:2002bv,Blaizot:2003br,Andersen:2004re}.   Historically, the $\Phi$-derivable approximation was first applied to QCD by Freedman and McLerran \cite{Freedman:1976ub}, who demonstrated that the thermodynamic potential $\Omega$ is gauge dependent beyond a given order in the coupling constant. The gauge parameter dependence appears at the same order in $\alpha_s$ as the series truncation when evaluated off the stationary point and at twice the order in $\alpha_s$ when evaluated at the stationary point \cite{Arrizabalaga:2002hn,Blaizot:1999ip,Andersen:2004re}. Despite this issue, this method had been used as the starting point for approximately self-consistent HTL resummation of the entropy \cite{Blaizot:1999ap,Blaizot:2000fc} and the pressure \cite{Peshier:2000hx}.

The problems encountered when applying the $\Phi$-derivable approximation to gauge theories motivated the use of alternative variational approximations. One such alternative, which in its simplest form involves a single variational parameter $m$, has been called optimized perturbation theory \cite{Stevenson:1981vj},  variational perturbation theory \cite{kleinert2009path,Sisakian:1994nn}, or the linear $\delta$ expansion \cite{Duncan:1988hw,Duncan:1992ba}. This strategy has been successfully used for the thermodynamics of the massless $\phi^4$ field theory up to four-loop order using ``screened perturbation theory'' \cite{Karsch:1997gj,Andersen:2000yj,Andersen:2001ez,Andersen:2008bz}, and spontaneously broken field theories at finite temperature \cite{Chiku:1998kd,Pinto:1999py,Chiku:2000eu,Kneur:2010yv}.  One impediment to applying such ideas to gauge theories was that one cannot simply introduce a scalar mass for the gluon without breaking gauge invariance.  The solution to this problem was introduced in Refs.~\cite{Andersen:1999sf} and \cite{Andersen:1999va} in which it was demonstrated that one could generalize the linear delta expansion by adding and subtracting the full gauge-invariant HTL effective Lagrangian \cite{Braaten:1991gm,Mrowczynski:2004kv}.

The resulting scheme was called hard-thermal-loop perturbation theory (HTLpt) ~\cite{Andersen:1999sf,Andersen:1999va}. The HTLpt method has been used to improve the convergence of weak coupling calculations of the free energy in QED \cite{Andersen:2009tw} and QCD up to three-loop order at finite temperature and chemical potential~\cite{Andersen:2009tc,Andersen:2010ct,Andersen:2010wu,Andersen:2011sf,Andersen:2011ug,Haque:2013sja,Haque:2014rua}.  When confronted with finite temperature and chemical potential lattice QCD data the HTLpt resummation scheme works remarkably well down to temperatures on the order of 200-300 MeV where the QCD coupling constant is on the order of $g_s \sim 2$ \cite{Haque:2013sja,Haque:2014rua,Ghiglieri:2020dpq}.  The method successfully describes all thermodynamic variables including second- and fourth-order quark susceptibilities.  Herein, we will take the first steps in applying this method to ${\mathcal N}=4$ SYM in the hope that a similar improvement in convergence can be achieved in this theory at intermediate couplings.  We will calculate the one- and two-loop HTLpt-resummed thermodynamic potential Additionally, in ${\mathcal N}=4$ SYM using the same method as was used to obtain the one- and two-loop QCD results in Refs.~\cite{Andersen:1999sf,Andersen:1999va,Andersen:2002ey,Andersen:2003zk}.  Importantly, in these papers it was demonstrated that it was possible to renormalize the resummed thermodynamic potential at two-loop order using only known vacuum and mass counterterms.  Herein we will demonstrate the same occurs in $\mathcal{N}=4$ SYM.  In this theory the NLO contributions include scalar and scalar-gluon, scalar-quark interactions compared to the QCD calculation, however these are relatively straightforward to include.  In  $\mathcal{N}=4$ SYM instead of having only gluon and quark thermal masses, $m_D$ and $m_q$, respectively, we will also have a thermal mass for the scalar particles, $M_D$.  Our results at one- and two-loop order are infinite series in $\lambda$ which, when trucated at ${\mathcal O}(\lambda^{3/2})$, reproduce the weak-coupling results obtained previously in Refs.~\cite{Fotopoulos:1998es,Kim:1999sg,VazquezMozo:1999ic}.  In order to make the calculation tractable, we expand the HTLpt scalarized sum-integrals in a power series in the three mass parameters $M_D$, $m_D$, and $m_q$ such that it includes terms that would naively contribute throughx ${\mathcal O}(\lambda^{5/2})$.  Our final results indicate that, in $\mathcal{N}=4$ SYM, NLO HTLpt provides a good approximation for the scaled entropy for couplings in the range $0 \leq \lambda \lesssim 2$.

We begin with a brief summary of HTLpt for $\mathcal{N}=4$ SYM in Sec.~\ref{theory}. In Sec.~\ref{potential}, we give the expressions for the one- and two-loop diagrams contributing to the SYM thermodynamic potential. In Sec.~\ref{reduce}, we reduce these diagrams to scalar sum-integrals. As mentioned in the prior paragraph, since it would be intractable to calculate the resulting sum-integrals analytically, in Sec.~\ref{expansion} we expand these expressions by treating $m_D$, $M_D$ and $m_q$ as $\mathcal{O}(\lambda^{1/2})$ and expanding the integrals in powers of $m_D/T$, $M_D/T$ and $m_q/T$, keeping all terms up to $\mathcal{O}(\lambda^{5/2})$. In Sec.~\ref{combine}, we combine the results obtained in Sec.~\ref{expansion} to obtain the complete expressions for the leading- (LO) and next-to-leading order (NLO) thermodynamic potentials. In Sec.~\ref{entropy}, we present our numerical results for the HTLpt-resummed LO and NLO scaled thermodynamic functions in $\mathcal{N}=4$ SYM and compare to prior results in the literature. For details concerning the transformation to Euclidean space and the sum-integrals necessary we refer the reader to the appendices of Refs.~\cite{Andersen:2002ey,Andersen:2003zk}.

\subsection*{Notation and conventions}

We use lower-case letters for Minkowski space four-vectors, e.g. $p$, and upper-case letters for Euclidean space four-vectors, e.g. $P$.  We use the mostly minus convention for the metric.

\section{HTLpt for $\mathcal{N}=4$ SYM}
\la{theory}

In $\mathcal{N}=4$ SYM theory all fields belong to the adjoint representation of the $SU(N_c)$ gauge group. For the fermionic fields, a massless two-component Weyl fermion $\psi$ in four dimensions can be converted to four-component Majorana fermions \cite{Quevedo:2010ui,bertolini2015lectures,Yamada:2006rx,DHoker:1999yni,Kovacs:1999fx}
\be
\psi \equiv  \begin{pmatrix} \psi_\alpha\\  \bar{\psi}^{\dot{\alpha}} \end{pmatrix}  \quad\quad \textrm{and} \quad\quad \bar{\psi} \equiv  \begin{pmatrix} \psi^\alpha & \bar{\psi}_{\dot{\alpha}} \end{pmatrix} ,
\ee
where $\alpha=1,2$ and the Weyl spinors satisfy $\bar{\psi}^{\dot{\alpha}}\equiv [\psi^\alpha]^\dagger$. The conjugate spinor $\bar{\psi}$ is not independent, but is related to $\psi$ via the Majorana condition $\psi=C\bar{\psi}$, where $C=\begin{pmatrix}\begin{smallmatrix} \epsilon_{\alpha\beta} & 0 \\ 0 & \epsilon^{\dot{\alpha}\dot{\beta}}\end{smallmatrix} \end{pmatrix}$ is the charge conjugation operator with $\epsilon_{02}=-\epsilon_{11}\equiv-1$. In the following, we will use the indices $i,j=1,2,3,4$ to enumerate the Majorana fermions and use $\psi_i$ to denote each bispinor.

The definition of gauge field is the same as QCD, and $A_\mu$ can be expanded as \mbox{$A_\mu=A_\mu^a t^a$}, with real coefficients $A_\mu^a$, and Hermitian color generators $t^a$ in the fundamental representation that satisfy
\ba\la{A}
 [t^a,t^b]=i f_{abc}t^c   \quad \textrm{and} \quad \textrm{Tr}(t^a t^b)=\frac{1}{2}\delta^{ab} \, ,
\ea
where $a,b=1, \cdots ,N_c^2-1$, the structure constants $f_{abc}$ are real and completely antisymmetric. The fermionic fields can similarly be expanded in the basis of color generators as $\psi_i=\psi_i^a t^a$. The coefficients $\psi_i^a$ are four-component Grassmann-valued spinors.

There are six independent real scalar fields which are represented by a multiplet
\ba\la{la}
 \Phi\equiv (X_1,Y_1,X_2,Y_2,X_3,Y_3) \, ,
\ea
where $X_{\texttt{p}}$ and $Y_{\texttt{q}}$ hermitian, with ${\texttt{p,q}}=1,2,3$. $X_{\texttt{p}}$ and $Y_{\texttt{q}}$ denote scalars and pseudoscalar fields, respectively. We will use a capital Latin index $A$ to denote components of vector $\Phi$. Therefore $\Phi_A$, $X_{\texttt{p}}$, and $Y_{\texttt{q}}$ can be expanded as $\Phi_A=\Phi_A^a t^a$, with $A=1, \cdots ,6$, and $X_{\texttt{p}}=X_{\texttt{p}}^at^a$, $Y_{\texttt{q}}=Y_{\texttt{q}}^at^a$.

The Lagrangian density that generates the perturbative expansion for $\mathcal{N}=4$ SYM theory in Minkowski-space can be expressed as
\ba\la{lag}
\mathcal{L}_{\textrm{SYM}}&=& \textrm{Tr}\bigg[{-}\frac{1}{2}G_{\mu\nu}^2+(D_\mu\Phi_A)^2+i\bar{\psi}_i {\displaystyle{\not} D}\psi_i-\frac{1}{2}g^2(i[\Phi_A,\Phi_B])^2 \nonumber \\
&& \hspace{1cm} - i g \bar{\psi}_i\big[\alpha_{ij}^{\texttt{p}} X_{\texttt{p}}+i \beta_{ij}^{\texttt{q}}\gamma_5Y_{\texttt{q}},\psi_j\big] \bigg] +\mathcal{L}_{\textrm{gf}}+\mathcal{L}_{\textrm{gh}}+\Delta\mathcal{L}_{\textrm{SYM}}\, ,
\ea
where the field strength tensor is $G_{\mu\nu}=\partial_\mu A_\nu-\partial_\nu A_\mu-ig[A_\mu,A_\nu]$, and $D_\nu=\partial_\nu-i g[A_\nu,\cdot]$ is the covariant derivation in the adjoint representation. $\alpha^{\texttt{p}}$ and $\beta^{\texttt{q}}$ are $4\times 4$ matrices that satisfy
\ba\la{lag1}
\{\alpha^{\texttt{p}},\alpha^{\texttt{q}}\}=-2\delta^{\texttt{pq}}, \quad \{\beta^{\texttt{p}},\beta^{\texttt{q}} \}=-2\delta^{\texttt{pq}}, \quad [\alpha^{\texttt{p}},\beta^{\texttt{q}}]=0 \, ,
\ea
and their explicit form can be given as
\ba\la{lag1x}
&&\alpha^1=\begin{pmatrix} 0& \sigma_1 \\ -\sigma_1 &0 \end{pmatrix},  \quad \quad \alpha^2=\begin{pmatrix} 0& -\sigma_3 \\ \sigma_3 &0 \end{pmatrix},  \quad \quad  \alpha^3=\begin{pmatrix} i\sigma_2& 0 \\ 0&i\sigma_2 \end{pmatrix},  \nonumber \\&& \beta^1=\begin{pmatrix} 0& i\sigma_2 \\ i\sigma_2 &0 \end{pmatrix},  \quad \quad  \beta^2=\begin{pmatrix} 0& \sigma_0 \\ -\sigma_0 &0 \end{pmatrix},  \quad \quad  \beta^3=\begin{pmatrix} -i\sigma_2&0 \\ 0 & i\sigma_2 \end{pmatrix} ,
\ea
where $\sigma_i$ with $i \in \{1,2,3\}$ are the $2\times 2$ Pauli matrices. And $\alpha$ and $\beta$ satisfies $\alpha_{ik}^{\texttt{p}}\alpha_{kj}^{\texttt{p}}=-3\delta_{ij}$ and $\beta_{ij}^{\texttt{q}}\beta_{ji}^{\texttt{p}}=-4\delta^{\texttt{pq}}$, with $\delta_{ii}=4$ for four Majorana fermions and $\delta^{\texttt{pp}}=3 $ for three scalars.

The ghost term $\mathcal{L}_{\textrm{gh}}$ depends on the choice of the gauge-fixing term $\mathcal{L}_{\textrm{gf}}$ and is the same as in QCD.  Here we work in general covariant gauge, giving
\ba\la{gf}
 \mathcal{L}_{\textrm{gf}}&=& -\frac{1}{\xi}\textrm{Tr}\big[(\partial^\mu A_\mu)^2 \big],  \nonumber \\\mathcal{L}_{\textrm{gh}}&=&-2\textrm{Tr}\big[\bar{\eta}\,\partial^\mu \! D_\mu \eta \big] ,
\ea
with $\xi$ being the gauge parameter.

In general, perturbative expansion in powers of $g$ in quantum field theory generates ultraviolet divergences. The renormalizability of perturbation theory guarantees that all divergences in physical quantities can be removed by the renormalization of masses and coupling constants. The coupling constant in $\mathcal{N}=4$ SYM theory is denoted as $\lambda=g^2N_c$.  Unlike QCD, $\mathcal{N}=4$ SYM theory does not run.

Similar to the case of QCD presented in Refs.~\cite{Andersen:2002ey,Andersen:2003zk}, HTLpt is also a reorganziation of the perturbation series for the SYM theory, and can be defined by introducing an expansion parameter $\delta$. The HTL ``shifted'' Lagrangian density can be written as
\ba\la{dla}
\mathcal{L}=(\mathcal{L}_{\textrm{SYM}}+\mathcal{L}_{\textrm{HTL}})|_{g\rightarrow\sqrt{\delta}g}+\Delta\mathcal{L}_{\textrm{HTL}} \, .
\ea
The HTL improvement term is
\ba\la{htl}
\mathcal{L}_{\textrm{HTL}}&=& -\frac{1}{2}(1-\delta)m_D^2\textrm{Tr}\bigg (G_{\mu \alpha} \bigg \langle\frac{y^\alpha y^\beta}{(y\cdot D)^2}\bigg \rangle_{\hat{\textbf{y}}} G_\beta^\mu \bigg )\nonumber
\\&&\hspace{1cm}
+ (1-\delta)i m_q^2 \textrm{Tr}\bigg (\bar{\psi}_j \gamma^\mu \bigg \langle \frac{y^\mu}{y\cdot D}   \bigg \rangle_{\hat{\textbf{y}}} \psi_j  \bigg ) \nonumber
\\&&\hspace{2cm}-(1-\delta) M_D^2 \textrm{Tr} (\Phi_A^2 ) \, ,
\ea
where $y^\mu=(1,\hat{\textbf{y}})$ is a light-like four vector defined in App.~\ref{fmr}, $j \in \{1  \ldots 4\}$ indexes the four Majorana fermions, $A \in \{ 1 \ldots 6\}$ indexes the scalar degrees of freedom, and $ \langle\cdots \rangle_{\hat{\textbf{y}}}$ represents the average over the direction of $\hat{\textbf{y}}$. The parameters $m_D$ and $M_D$ are the electric screening masses for the gauge field and the adjoint scalar field, respectively. The parameter $m_q$ can be seen as the induced finite temperature quark mass. We note that, if we set $\delta=1$, the Lagrangian above \eqref{dla} reduces to the vacuum $\mathcal{N}=4$ SYM Lagrangian (\ref{lag}).   HTLpt is defined by treating $\delta$ as a formal expansion parameter, expanding around $\delta=0$ to a fixed order, and then setting $\delta=1$.  In the limit that this expansion is taken to all orders, one reproduces the QCD result by construction, however, the loop expansion is now shifted to be around the high-temperature minimum of the effective action, resulting in a reorganization of the perturbation series which has better convergence than the naive expansion loop expansion around the $T=0$ vacuum.  In addition, this reorganization eliminates all infrared divergences associated with the electric sector of the theory.

The HTLpt reorganization generates new ultraviolet (UV) divergences and, due to the renormalizability of perturbation theory, the ultraviolet divergences are constrained to have a form that can be canceled by the counterterm Lagrangian $\Delta\mathcal{L}_{\textrm{HTL}}$. References \cite{Andersen:2002ey,Andersen:2003zk} demonstrated that at two-loop order the thermodynamic potential can be renormalized using a simple counterterm Lagrangian $\Delta \mathcal{L}_{\textrm{HTL}}$ containing vacuum and mass counterterms. Although the general structure of the ultraviolet divergences is unknown, it has been demonstrated that one can renormalize the next-to-leading order HTLpt thermodynamic potential through three-loop order using only vacuum, gluon thermal mass, quark thermal  mass, and coupling constant counterterms~\cite{Andersen:2011sf,Haque:2014rua}. In this paper, we demonstrate that the same method can be used for $\mathcal{N}=4$ SYM and we compute the vacuum and screening mass counterterms necessary.

We find that the vacuum counterterm $\Delta_0 \mathcal{E}_0$, which is the leading order counterterm in the $\delta$ expansion of the vacuum energy $\mathcal{E}_0$, can be obtained by calculating the free energy to leading order in $\delta$. In Sec.~\ref{leading}, we show that $\Delta_1 \mathcal{E}_0$ can be obtained by expanding $\Delta \mathcal{E}_0$ to linear order in $\delta$. As a result, the counterterm $\Delta \mathcal{E}_0$ has the form
\ba\la{dc0}
 \Delta \mathcal{E}_0=\bigg(\frac{d_A}{128\pi^2 \epsilon} +O(\delta \lambda) \bigg)(1-\delta)^2m_D^4+\bigg(\frac{3d_A}{32\pi^2 \epsilon}+O(\delta \lambda)\bigg)(1-\delta)^2 M_D^4 \,.
\ea
To calculate the NLO free energy we need to expand to order $\delta$ and we will need the counterterms $\Delta\mathcal{E}_0$, $\Delta m_D^2$, $\Delta m_q^2$, and $\Delta M_D^2$ to order $\delta$ in order to cancel the UV divergences. We find that in order to remove the divergences to two-loop order, the mass counterterms should have the form
\ba\la{dmass}
\Delta m_D^2&=&\bigg(\frac{1}{16\pi^2\epsilon}\delta\lambda +O(\delta^2 \lambda^2) \bigg)(1-\delta) m_D^2  \, ,  \nonumber \\
\Delta M_D^2&=&\bigg(\frac{3}{8\pi^2 \epsilon}\delta\lambda + O(\delta^2 \lambda^2) \bigg)(1-\delta)  M_D^2 \,,   \nonumber \\
\Delta m_q^2&=&\bigg(-\frac{1}{\pi^2 \epsilon}\delta\lambda + O(\delta^2 \lambda^2) \bigg)(1-\delta) m_q^2 \,\,.
\ea

In the $\mathcal{N}=4$ SYM theory, we will use the same method as in QCD to calculate physical observables in HTLpt, namely expanding the path-integral in powers of $\delta$, truncating at some specified order, and then setting $\delta=1$. The results of the physical observables will depend on $m_D$, $M_D$, and $m_q$ for any truncation of the expansion in $\delta$, and some prescription is required to determine $m_D$, $M_D$, and $m_q$ as a function of $\lambda$. In this work, we will follow the two-loop HTLpt QCD prescription and determine them by minimizing the free energy. If we use $\Omega_N(T,\lambda, m_D, M_D, m_q,  \delta)$ to represent the thermodynamic potential expanded to $N$-th order in $\delta$, then our full variational prescription is
\ba\la{gap}
\frac{\partial}{\partial m_D}\Omega_N(T, \lambda, m_D, M_D, m_q, \delta=1)&=&0,      \nonumber \\
\frac{\partial}{\partial M_D}\Omega_N(T, \lambda, m_D, M_D, m_q, \delta=1)&=&0,     \nonumber \\
\frac{\partial}{\partial m_q}\Omega_N(T, \lambda, m_D, M_D, m_q, \delta=1)&=&0 \,.
\ea
We will call Eqs.~(\ref{gap}) the \textit{gap equations}. The free energy is obtained by evaluating the thermodynamic potential at the solution to the gap equations. Other thermodynamic functions can then be obtained by taking appropriate derivatives of free energy with respect to $T$.

\section{Next-to-leading order thermodynamic potential}
\la{potential}

In the imaginary-time formalism, Minkoswski energies have discrete imaginary values $p_0=i(2\pi nT)$, and the integrals over Minkowski space should be replaced by Euclidean sum integrals. There are two ways to do this which have been discussed in Refs.~\cite{Andersen:2002ey,Andersen:2003zk}. One is transforming the Feynman rules in Minkowski space given in App.~\ref{fmr} into the form in Euclidean space firstly, then calculating the free energy. The other way is using the Feynman rules in Minkowski space to get the forms of free energy, after reducing these forms, transforming it into the form in Euclidean space. Results from the two methods must be the same.

\begin{figure}[t!]
\centering
\includegraphics[width=0.58\textwidth]{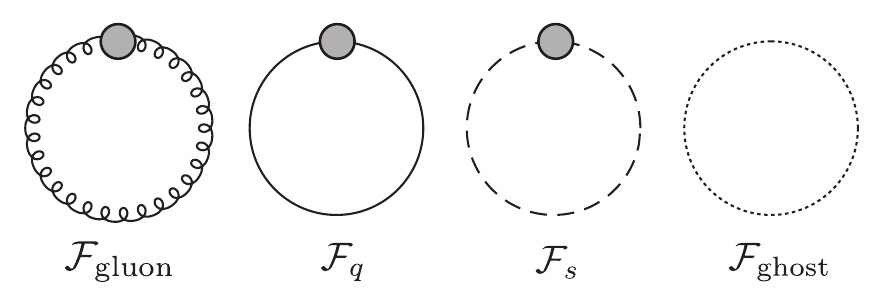}
\caption{\label{1-o}
One loop Feynman diagrams for $\mathcal{N}=4$ SYM theory in HTLpt. Dashed lines indicate a scalar field and dotted lines indicate a ghost field. Shaded circles indicate HTL-dressed propagators.}
\end{figure}

The HTL perturbative thermodynamic potential at next-to-leading order in $\mathcal{N}=4$ SYM can be expressed as
\ba\la{nel}
\Omega_{\textrm{NLO}}= \Omega_{\textrm{LO}}+\Omega_{\textrm{2-loop}}+\Omega_{\textrm{HTL}}+ \Delta\Omega_{\textrm{NLO}}     \, ,
\ea
where $\Omega_{\textrm{LO}}$ is the leading order thermodynamic potential, ${\mathcal O}(\delta^0)$, which includes the one-loop graphs shown in Fig.~\ref{1-o} and the LO vacuum renormalization counterterm.  We first discuss the contributions at this order.

\subsection{LO thermodynamic potential}
\la{oneloopomega}

In $\textrm{SU}(N_c)$ gauge theory with massless particles, $\Omega_{\textrm{LO}}$ can be expressed as
\ba\la{onel}
\Omega_{\textrm{LO}}=d_A \mathcal{F}_g +d_F\mathcal{F}_q+d_S\mathcal{F}_s+\Delta_0\mathcal{E}_0 \, ,
\ea
where $d_A=N_c^2-1$ is the dimension of the adjoint representation. There are four independent Majorana fermions in the adjoint representation, $d_F=4d_A$, and $d_S=6d_A$ for the six scalars.

There are $D=d+1$ polarization state for gluons, where $d$ is the number of spatial dimensions.  After canceling the two unphysical states using the ghost contribution, we obtain the HTL one-loop free energy of each of the color states of the gluon
\ba\la{oneg}
\mathcal{F}_g = \mathcal{F}_{\rm gluon} + \mathcal{F}_{\rm ghost} = -\frac{1}{2}\sumint_P \{(d-1)\log[-\Delta_T(P)]+\log\Delta_L(P) \}\,.
\ea
The transverse and longitudinal HTL propagators $\Delta_T(P)$ and $\Delta_L(P)$ are the HTL gluon propagator (\ref{tslo}) in Euclidean space
\ba\la{prog2x}
\Delta_T(P)&=&\frac{-1}{P^2+\Pi_T(P)},   \nonumber \\
\Delta_L(P)&=&\frac{1}{p^2+\Pi_L(P)}   \,.
\ea
The result above is the same as in QCD.  The only difference is the definition of $m_D^2$ in the gluon propagator which now contains contributions from gluon, fermion, and scalar loops as detailed in Appendix \ref{gsfn}.

Since the Majorana fermion is its own antiparticle, the fermionic contribution is reduced by a factor of two when comparing QCD and $\mathcal{N}=4$ SYM.  Our definition of $m_q^2$ is presented in Appendix \ref{qsfn}.   The one-loop fermionic free energy  is
\ba\la{oneq}
\mathcal{F}_q=-\frac{1}{2} \, \sumint_{\{P\}}\log \textrm{det}[{\displaystyle{\not} P}-\Sigma(P) ] \, ,
\ea
where $\Sigma(P)$ is the HTL fermion self-energy \eqref{sfq} in Euclidean space. The scalar one-loop free energy is simply
\ba\la{ones}
\mathcal{F}_s=\frac{1}{2} \, \sumint_{P}\log[-\Delta_s^{-1}(P)] \, ,
\ea
where $\Delta^{-1}_s(P)$ is the inverse scalar propagator which is given in Eq.~(\ref{pros2x}).   Finally, we note that the leading order counterterm $\Delta_0 \mathcal{E}_0$ cancels the divergent terms of the one-loop thermodynamic potential in $\mathcal{N}=4$ SYM theory.

\subsection{NLO thermodynamic potential}
\la{twoloopomega}

\begin{figure}[t!]
\centering
\includegraphics[width=0.75\textwidth]{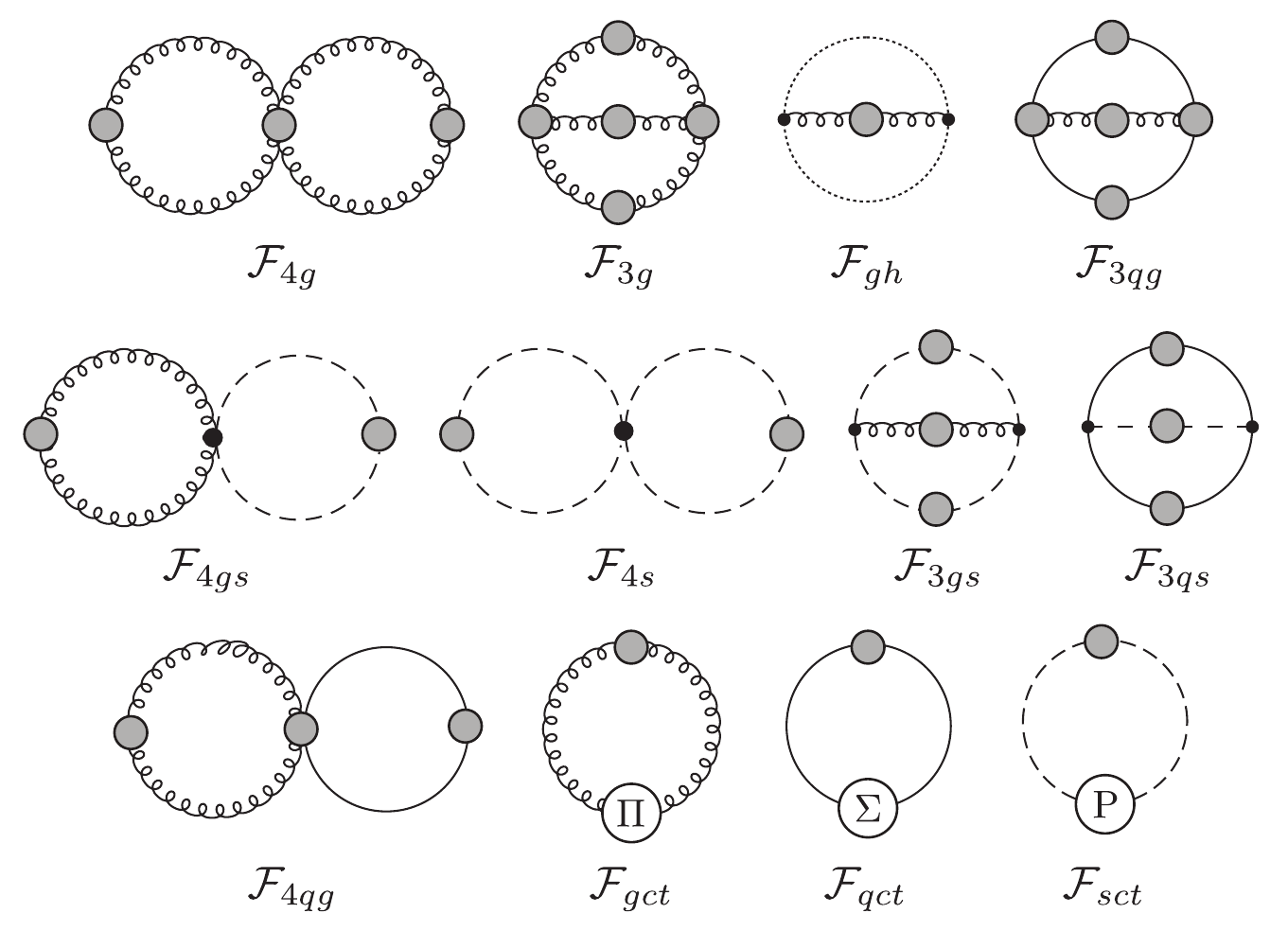}
\caption{\label{2-o}
Two loop Feynman diagrams for $\mathcal{N}=4$ SYM theory in HTLpt. Dashed lines indicate a scalar field and dotted lines indicate a ghost field. Shaded circles indicate HTL-dressed propagators and vertices.}
\end{figure}

In Eq.~(\ref{nel}) $\Omega_{\textrm{2-loop}}$ corresponds to the two-loop contributions shown in Fig.~\ref{2-o}.  It can be expressed as
\ba\la{nelx}
\Omega_{\textrm{2-loop}}= d_A \lambda \big[ \mathcal{F}_{3g}+\mathcal{F}_{4g} +\mathcal{F}_{gh} +\mathcal{F}_{4s}+\mathcal{F}_{3gs} +\mathcal{F}_{4gs}  + \mathcal{F}_{3qg}+\mathcal{F}_{4qg} +\mathcal{F}_{3qs}  \big] ,
\ea
where $\lambda = g^2 N_c$ is the 't Hooft coupling constant.

The gluon propagator, the three-gluon vertex, the four-gluon vertex, and the gluon-ghost vertex are the same as in QCD up to the expression for the Debye mass $m_D$.\footnote{See \ref{gsfn} for proof of this statement.}  As a result, the purely gluonic and glue-ghost graphs given by $\mathcal{F}_{3g}$, $\mathcal{F}_{4g}$ , and $\mathcal{F}_{gh}$ are, respectively,
\ba\la{f3g}
\mathcal{F}_{3g}&=&\frac{1}{12} \, \sumint_{PQ} \Gamma^{\mu\lambda\rho}(P,Q,R)\Gamma^{\nu\sigma\tau}(P,Q,R)\Delta^{\mu\nu}(P)\Delta^{\lambda\sigma}(Q)\Delta^{\rho\tau}(R),    \nonumber \\
\mathcal{F}_{4g}&=&\frac{1}{8} \, \sumint_{PQ}\Gamma^{\mu\nu,\lambda\sigma}(P,-P,Q,-Q)\Delta^{\mu\nu}(P)\Delta^{\lambda\sigma}(P),    \nonumber \\
\mathcal{F}_{gh}&=&\frac{1}{2} \, \sumint_{PQ}\frac{1}{Q^2}\frac{1}{R^2}Q^\mu R^\nu\Delta^{\mu\nu}(P)    \, ,
\ea
where $R+P+Q=0$. Contributions come from the two-loop diagrams with four scalar vertex, scalar-gluon vertex and scalar-gluon four vertex are respectively
\ba\la{f4s}
\mathcal{F}_{4s}&=&\frac{15}{2} \, \sumint_{PQ}\Delta_s(P)\Delta_s(Q),    \nonumber \\
\mathcal{F}_{3gs}&=&\frac{3}{2} \, \sumint_{PQ}\Delta_s(R)\Delta_s(Q)\Delta_{\mu\nu}(P)(R+Q)_\mu(R+Q)_\nu,   \nonumber \\
\mathcal{F}_{4gs}&=&3\,\sumint_{PQ}\Delta_s(Q)\Delta_{\mu \nu}(P)\delta_{\mu\nu}      \, ,
\ea
where $R+P-Q=0$. The contributions $\mathcal{F}_{3qg}$ and $\mathcal{F}_{4qg}$ involve only quarks and gluons and, since the Majorana fermion is its own antiparticle, their symmetry factor is $1/4$ instead of $1/2$ in QCD.  Additionally, there are four Majorana fermions in $\mathcal{N}=4$ SYM, so that $\mathcal{F}_{3qg}$ and $\mathcal{F}_{4qg}$ are 2 times the result obtained in QCD.  As a result, we can substitute $s_F$ and $d_F$ in Ref.~\cite{Andersen:2003zk} to $2N_c$ and $2d_A$, respectively, to obtain the $\mathcal{N}=4$ SYM result.  After this adjustment, the only other change required is to use the $\mathcal{N}=4$ SYM definitions of $m_D^2$ and $m_q^2$.  Based on the results contained in Ref.~\cite{Andersen:2003zk} one obtains
\ba\la{fgq}
\mathcal{F}_{3qg}&=&-\sumint_{P\{Q\}}\Delta_{\mu\nu}(P)\textrm{Tr}\big[\Gamma^\mu(P,Q,R)S(R)\Gamma^\nu(P,Q,R)S(Q)   \big],    \nonumber \\
\mathcal{F}_{4qg}&=&- \sumint_{P\{Q\}}\Delta_{\mu\nu}(P)\textrm{Tr}\big[\Gamma^{\mu\nu}(P,-P,Q,Q)S(Q)  \big] .
\ea
The momentum conservation is $R+P-Q=0$. One can also obtain these expressions using the Feynman rules contained in App.~\ref{fmr}.

The final new graph, the quark-scalar diagram $\mathcal{F}_{3qs}$, can be split into two parts, one coming from the quark-scalar vertex, and the other coming from the quark-pseudoscalar vertex.  Using the Feynman rules in App.~\ref{fmr}, one finds that their contributions are the same. As a consequence, $\mathcal{F}_{3qs}$ can be written as
\ba\la{fqs}
\mathcal{F}_{3qs}&=&-6 \, \sumint_{P\{Q\}}\textrm{Tr}\big[ S(R)S(Q) \big]\Delta_s(P)  \, ,
\ea
where $R+P-Q=0$.

The contribution $\Omega_{\textrm{HTL}}$ in Eq.~(\ref{nel}) is the sum of the gluon, quark and scalar HTL counterterms shown in Fig.~\ref{2-o}.  These enter in order to subtract contributions at lower loop orders and guarantee that naive perturbative results are recovered order by order if the expressions are truncated in $\lambda$.  They can be expressed as
\ba\la{nelx1}
\Omega_{\textrm{HTL}} = d_A \mathcal{F}_{gct} +d_F\mathcal{F}_{qct}  +d_S\mathcal{F}_{sct} \,.
\ea
There are two ways to get these three contributions, one is using the Feynman rules in Appendix \ref{fmr}, the other one is substituting $m_D^2\rightarrow (1-\delta)m_D^2$, $M_D^2\rightarrow (1-\delta)M_D^2$ and $m_q^2\rightarrow (1-\delta)m_q^2$ in the one-loop expressions for $\mathcal{F}_g$+$\mathcal{F}_{\rm ghost}$, $\mathcal{F}_s$, and $\mathcal{F}_q$ and expanding them to linear order in $\delta$. In terms of the first method, the contribution from the HTL gluon counterterm diagram is
\ba\la{gct}
\mathcal{F}_{gct}&=&\frac{1}{2}\sumint_{P}\Pi^{\mu\nu}(P)\Delta^{\mu\nu}(P)  \nonumber \\   &=&  \frac{1}{2}\sumint_{P}\big[(d-1)\Pi_T(P)\Delta_T(P)- \Pi_L(P)\Delta_L(P)   \big]\,.
\ea
It is the same as in QCD up to the definition of $m_D^2$. The contribution from the HTL scalar counterterm diagram is
\ba\la{sct}
\mathcal{F}_{sct}&=&\frac{1}{2}\sumint_{P}\Delta_s(P)\mathcal{P}_{aa}^{AA}(P) \, ,
\ea
where $\mathcal{P}_{aa}^{AA}(P)=M_D^2$ is referred in Appendix \ref{ssfn}. The contribution from the HTL quark counterterm diagram is
\ba\la{qct}
\mathcal{F}_{qct}&=&-\frac{1}{2}\sumint_{\{P\}}\textrm{Tr}\big[\Sigma(P)S(P)\big]\,.
\ea
Compared to QCD this is different by one half due to the Majorana nature of the SYM fermions.  As usual, the quark mass should be adjusted to the SYM case.

Since HTL perturbation theory is renormalizable, the ultraviolet divergences of free energy at any order in $\delta$ can be cancelled by $\mathcal{E}$, $m_D^2$, $m_q^2$, and $M_D^2$ and the coupling constant $\lambda$. $\Delta\Omega_{\rm NLO}$ in Eq.~(\ref{nel}) is the renormalization contribution at first order in $\delta$ is used to cancel the next-to-leading order divergences. It can be expressed as
\ba\la{nelx2}
\Delta\Omega_{\rm NLO} = \Delta_1 \mathcal{E}_0+ \big(\Delta_1m_D^2 +\Delta_1m_q^2+\Delta_1M_D^2\big)\frac{\partial}{\partial M_D^2} \Omega_{\textrm{LO}}  \, ,
\ea
where $\Delta_1 \mathcal{E}_0$, $\Delta_1m_D^2$, $\Delta_1m_q^2$, and $\Delta_1M_D^2$ are the terms of order $\delta$ in the vacuum energy \eqref{dc0} and mass counterterms \eqref{dmass}.  The first term $\Delta_1 \mathcal{E}_0$ can be obtained simply by expanding $\Delta_0 \mathcal{E}_0$ to first order in $\delta$. The second term in (\ref{nelx2}) is slightly different from the QCD result in Refs.~\cite{Andersen:2002ey,Andersen:2003zk}.  This is because this term must be used to cancel the divergences of two-loop self energy. As we can see in (\ref{2op}), there are two mixed term $M_D m_D^2$ and $M_D m_q^2$ which comes from the $(hs)$ contribution of $\mathcal{F}_{4gs}$ and $\mathcal{F}_{3qs}$ respectively. There are many ways to construct the mass renormalization form which is corresponding to the second part of eq.(\ref{nelx2}), but the only way to use one set of three counterterms $\Delta_1m_D^2$, $\Delta_1m_q^2$, and $\Delta_1M_D^2$ is the form we have shown above.

In this work, we calculate the thermodynamic potential as an expansion in powers of $m_D/T$, $m_q/T$, and $M_D/T$ to order $g^5$. We will show that, at order $\delta$, all divergences in the two-loop thermodynamic potential plus HTL counterterms can be removed by these vacuum and mass counterterms.  This means the method used in QCD can also be used in $\mathcal{N}=4$ SYM theory.  This provides nontrivial evidence for the renormalizability of HTLpt at order $\delta$ in $\mathcal{N}=4$ SYM.

\section{Reduction to scalar sum-integrals }
\la{reduce}

Since we can make use of prior QCD results, we only need to calculate $\mathcal{F}_{s}$, $\mathcal{F}_{sct}$, $\mathcal{F}_{4s}, \mathcal{F}_{3gs}$, $\mathcal{F}_{4gs}$, $\mathcal{F}_{3qs}$, and the HTL counterterms contributing to Eq.~\eqref{nelx1}. The first step to calculate the new SYM contributions in Figs.~\ref{1-o} and \ref{2-o} is to reduce the sum of these diagrams to scalar sum-integrals. In Euclidean space, by substituting $p_0$ to $i P_0$ the scalar propagator can be written as
\ba\la{pros1x}
\Delta_s(P)=\frac{-1}{P^2+M_D^2} \,,
\ea
so its inverse is
\ba\la{pros2x}
\Delta^{-1}_s(P)=-(P^2+M_D^2) \,.
\ea
The leading-order scalar contribution can be written as
\ba\la{ones1}
\mathcal{F}_s=\frac{1}{2}\sumint_{P}\log[P^2+M_D^2]\,.
\ea
The HTL scalar counterterm can be written as
\ba\la{sct1}
\mathcal{F}_{sct}&=&-\frac{1}{2}\sumint_{P}\frac{M_D^2}{P^2+M_D^2}\,.
\ea

We proceed to simplify the sum of formulas in Eq.~(\ref{f4s}) in general covariant gauge parameterized by $\xi$. Using Eqs.~(\ref{pros1x}) and (\ref{prog7}), we obtain
\ba\la{f4s1}
\mathcal{F}_{4s+3gs+4gs}&=&\frac{15}{2}\,\sumint_{PQ}\Delta_s(P)\Delta_s(Q) \nonumber \\
&&+\frac{3}{2}\,\sumint_{PQ}\Delta_s(R)\Delta_s(Q)\bigg\{\Delta_T(P)\bigg[ 2R^2+2Q^2-P^2-\frac{(Q^2-R^2)^2}{P^2} \bigg]\nonumber \\
&&+\Delta_X(P)\bigg[\frac{2}{P^2}(Q^2-R^2)(Q^2-R^2+r^2-q^2)-(2R^2+2Q^2-P^2)\nonumber \\
&&+(2r^2+2q^2-p^2)-\frac{P_0^2}{P^4}(Q^2-R^2)^2   \bigg]-\frac{\xi}{P^4}(Q^2-R^2)^2  \bigg\}  \nonumber \\
&&+3 \, \sumint_{PQ} \Delta_s(Q)\bigg[ d \Delta_T(P)-\frac{p^2}{P^2}\Delta_X(P)-\frac{\xi}{P^2} \bigg]    \,.
\ea
There are two terms which depend on $\xi$ in (\ref{f4s1}), however, using \mbox{$P+R-Q=0$}, one finds that they cancel each other, so that the sum of these contributions is gauge independent. Similarly, the results for $\mathcal{F}_{3g+4g+gh}$ and $\mathcal{F}_{3qg+4qg}$ are independent of gauge parameter $\xi$ as shown in prior QCD calculations.  Therefore, we have verified explicitly that the NLO HTLpt resummed thermodynamic potential in $\mathcal{N}=4$ SYM is gauge independent.

Finally, we simplify Eq.~(\ref{fqs}) to
\ba\la{fqs1}
\mathcal{F}_{3qs}&=&-24\,\sumint_{P\{Q\}}\frac{\mathcal{A}_0(R)\mathcal{A}_0(Q) - \mathcal{A}_s(R)\mathcal{A}_s(Q)\hat{\textbf{r}}\cdot \hat{\textbf{q}}}{[\mathcal{A}^2_0(R)-\mathcal{A}^2_s(R)][\mathcal{A}^2_0(Q)-\mathcal{A}^2_s(Q)]} \Delta_s(P) \, ,
\ea
where in Euclidean space
\ba\la{proq2e}
\mathcal{A}_0(P)&=&i P_0-\frac{m_q^2}{i P_0}\mathcal{T}_P \, ,     \nonumber \\
\mathcal{A}_s(P)&=&p+\frac{m_q^2}{p} \big[1-\mathcal{T}_P  \big] ,
\ea
with ${\mathcal T}_P$ is the angular average $\mathcal{T}^{00}(p,-p)$ in Euclidean space, can be expressed as
\be\la{t00x1}
\mathcal{T}_P=\bigg\langle\frac{P_0^2}{P_0^2+p^2c^2}   \bigg\rangle_c \,,
\ee
where the angular brackets denote an average over $c$ defined by
\be\la{t00x2}
\big\langle f(c)\big\rangle_c \equiv \omega(\epsilon) \int^1_0 dc (1-c^2)^{-\epsilon}f(c) \, ,
\ee
where $ \omega(\epsilon)$ is given in (\ref{t000}).

\section{High temperature expansion }
\la{expansion}

Having reduced $\mathcal{F}_s$, $\mathcal{F}_{4s+3gs+4gs}$, $\mathcal{F}_{3qs}$, and the HTL counterterm $\mathcal{F}_{sct}$ to scalar sum-integrals, we will now evaluate these sum-integrals approximately by expanding them in powers of $m_D/T$, $m_q/T$, and $M_D/T$. We will keep terms that contribute through ${\mathcal O}(g^5)$ if $m_D, m_q$ and $M_D$ are taken to be of order $g$ at leading-order.  Additionally, each of these terms can be divided into contributions from hard and soft momentum, so we will proceed to calculate their hard and soft contributions, respectively.  In some cases, the results presented here were obtained in previous one- and two-loop QCD HTLpt papers \cite{Andersen:1999va,Andersen:2002ey,Andersen:2003zk}.  When converting the prior QCD graphs involving quarks, as mentioned previously, one has to take into account that the four SYM quarks are Majorana fermions.  Here we list results for all contributions to the ${\mathcal N}=4$ SYM Feynman graphs and counterterms for completeness and ease of reference. In all cases, we use the integral and sum-integral formulas from Refs.~\cite{Andersen:2002ey,Andersen:2003zk} to obtain explicit expressions.

\subsection{One-loop sum-integrals }

The one-loop sum-integrals include the leading gluon, quark, and scalar contributions (\ref{oneg}), (\ref{oneq}), and (\ref{ones}) along with their corresponding counterterms (\ref{gct}), (\ref{qct}), and (\ref{sct}). In order to include all terms through ${\mathcal O}(g^5)$, we need to expand the one-loop contribution to order $m_D^4$, $m_q^4$, and $M_D^4$.

\subsubsection{ Hard contributions }

The hard contribution from the gluon free energy (\ref{oneg}) is \cite{Andersen:2002ey}
\ba\la{gh}
\mathcal{F}_g^{(h)}&=& -\frac{\pi^2}{45}T^4+\frac{1}{24}\bigg[ 1+ \bigg(2+2\frac{\zeta^{'}(-1)}{\zeta(-1)}   \bigg)\epsilon   \bigg]\bigg( \frac{\mu}{4\pi T}   \bigg)^{2\epsilon} m_D^2 T^2\nonumber \\
&& - \frac{1}{128\pi^2}\bigg(\frac{1}{\epsilon} -7+2\gamma+\frac{2\pi^2}{3}  \bigg)\bigg( \frac{\mu}{4\pi T}   \bigg)^{2\epsilon}m_D^4   \,.
\ea
The hard contribution from the gluon HTL counterterm (\ref{gct}) is \cite{Andersen:2002ey}
\ba\la{gcth}
\mathcal{F}_{gct}^{(h)}&=& -\frac{1}{24}m_D^2 T^2+\frac{1}{64\pi^2}\bigg( \frac{1}{\epsilon}-7+2\gamma+\frac{2\pi^2}{3}    \bigg)\bigg( \frac{\mu}{4\pi T}   \bigg)^{2\epsilon}m_D^4   \,.
\ea

The hard contribution from the quark free energy (\ref{oneq}) is
\ba\la{qh}
\mathcal{F}_q^{(h)}&=& -\frac{7\pi^2}{360}T^4+ \frac{1}{12}\bigg [1+\bigg (2-2\log 2+2\frac{\zeta^{'}(-1)}{\zeta(-1)} \bigg)\epsilon       \bigg] \bigg( \frac{\mu}{4\pi T}   \bigg)^{2\epsilon} m_q^2 T^2  \nonumber \\&+&\frac{1}{24\pi^2}\big( \pi^2-6    \big)m_q^4 \,.
\ea
The hard contribution from the quark HTL counterterm (\ref{qct}) is
\ba\la{qcth}
\mathcal{F}_{qct}^{(h)}&=& -\frac{1}{12}m_q^2T^2-\frac{1}{12\pi^2}\big(\pi^2-6   \big)m_q^4 \,.
\ea

Since scalars are bosons, the sum-integrals in (\ref{ones}) are the same those used for gluons.  After expansion,  we obtain the hard contribution to the LO scalar free energy
\ba\la{sh1}
\mathcal{F}_s^{(h)}=\frac{1}{2}\sumint_{P}\log P^2+\frac{1}{2}M_D^2\sumint_{P}\frac{1}{P^2}-\frac{1}{4}M_D^4 \sumint_{P}\frac{1}{P^4}  \, .
\ea
Using the results for sum-integrals contained in the Appendixes B and C of Refs.~\cite{Andersen:2002ey,Andersen:2003zk}, Eq.~(\ref{sh1}) reduces to
\ba\la{sh}
\mathcal{F}_s^{(h)}&=& -\frac{1}{90}\pi^2T^4+\frac{1}{24}\bigg [ 1+\bigg(2+ 2\frac{\zeta^{'}(-1)}{\zeta(-1)}  \bigg )\epsilon    \bigg]\bigg( \frac{\mu}{4\pi T}   \bigg)^{2\epsilon} M_D^2T^2 \nonumber \\
&&-\frac{1}{64\pi^2}\bigg[ \frac{1}{\epsilon} +2\gamma  \bigg]\bigg( \frac{\mu}{4\pi T}   \bigg)^{2\epsilon}M_D^4 \,.
\ea
The scalar HTL counterterm is given in (\ref{sct}), after expansion, we get
\ba\la{scth1}
\mathcal{F}_{sct}^{(h)}=-\frac{1}{2}M_D^2\sumint_{P}\frac{1}{P^2}+\frac{1}{2}M_D^4\sumint_{P}\frac{1}{P^4}    \,,
\ea
which can be reduced to
\ba\la{scth}
\mathcal{F}_{sct}^{(h)}=-\frac{1}{24}M_D^2T^2+\frac{1}{32\pi^2}\bigg (\frac{1}{\epsilon}+2\gamma    \bigg)\bigg( \frac{\mu}{4\pi T}   \bigg)^{2\epsilon}M_D^4   \,.
\ea
Note that the first terms in (\ref{gh}), (\ref{qh}) and (\ref{sh}) cancel the order-$\epsilon^0$ term in the coefficient of mass squared in (\ref{gcth}), (\ref{qcth}) and (\ref{scth}), respectively.

\subsubsection{Soft contributions }

The {\em soft} contributions to the thermodynamical potential come from the $n=0$ Matsubara mode ($P_0=0$) in the resulting bosonic sum-integrals. For fermions, since $P_0=(2n+1)\pi T\neq 0$ for integer $n$, the quark momentum is always {\em hard}; therefore, quarks do not have a soft contribution. For gluons, in the soft limit $P \rightarrow (0,{\bf p})$, the HTL gluon self-energy functions reduce to $\Pi_T(P)=0 $ and $\Pi_L(P)=m_D^2$.  For scalars, in this limit the propagator reduces to $\Delta_s(P)=-1/(p^2+M_D^2)$ where, here $p^2 = {\bf p}^2$.

The soft contribution to the gluon free energy (\ref{oneg}) is
\ba\la{gs}
\mathcal{F}_g^{(s)}= -\frac{1}{12\pi}\bigg(1+\frac{8}{3}\epsilon   \bigg)\bigg(  \frac{\mu}{2m_D}   \bigg)^{2\epsilon}m_D^3 T\,.
\ea
The soft contribution from the gluon HTL counterterm (\ref{gct}) is
\ba\la{gcts}
\mathcal{F}_{gct}^{(s)}=\frac{1}{8\pi}m_D^3 T \,.
\ea
The soft contribution from scalar free energy (\ref{ones}) is
\ba\la{ss1}
\mathcal{F}_s^{(s)}=\frac{1}{2}T\int_{\textbf{p}}\log\big(p^2+M_D^2   \big)  \,,
\ea
which can be reduced to
\ba\la{ss}
 \mathcal{F}_s^{(s)}= -\frac{1}{12\pi}\bigg(1+\frac{8}{3}\epsilon   \bigg)\bigg(  \frac{\mu}{2M_D}   \bigg)^{2\epsilon}M_D^3 T \,.
\ea
The soft contribution from the scalar HTL counterterm (\ref{sct}) is
\ba\la{scts1}
\mathcal{F}_{sct}^{(s)}=-\frac{1}{2}M_D^2T\int_{\textbf{p}}\frac{1}{p^2+M_D^2}   \,,
\ea
then it can be reduced as
\ba\la{scts}
\mathcal{F}_{sct}^{(s)}=\frac{1}{8\pi}M_D^3 T  \,.
\ea

\subsection{Two-loop sum-integrals}

Since the two-loop sum-integrals have an explicit factor of $\lambda$, we only need to expand these sum-integrals to order $m_D^3/T^3$, $M_D^3/T^3$, $m_D m_q^2/T^3$, $M_D m_q^2/T^3$, $m_D^2 M_D/T^3$, and $M_D^2 m_D/T^3$ to include all terms through $\lambda^{5/2}$. Since these integrals involve two momentum integrations we will expand contributions from hard loop momentum and soft loop momentum for each momentum integral. For bosons, this gives three contributions which we will denote as $(hh)$, $(hs)$ and $(ss)$. For fermions, since their momentum is always hard, there will be only two regions $(hh)$ and $(hs)$. In the $(hh)$ region, all three momentum are hard $p \sim T$, while in the $(ss)$ region, all the three momentum are soft, $p \sim gT$. In the $(hs)$ region, two of the three momenta are hard and the other is soft.

\subsubsection{ Contributions from the $(hh)$ region}

In the $(hh)$ region, the self energies are suppressed by $m_D^2/T^2$, $M_D^2/T^2$ and $m_q^2/T^2$, so we can expand in powers of $\Pi_T$, $\Pi_L$, $\Sigma$, and $M_D^2$.

The $(hh)$ contribution from gluon self energy (\ref{f3g}) is \cite{Andersen:2002ey}
\ba\la{ghh}
\mathcal{F}_{3g+4g+gh}^{(hh)}= \frac{1}{144} T^4-\frac{7}{1152\pi^2}\bigg( \frac{1}{\epsilon}+4.6216  \bigg)\bigg( \frac{\mu}{4\pi T}   \bigg)^{4\epsilon} m_D^2T^2\,.
\ea
The $(hh)$ contribution from quark self energy (\ref{fgq}) is \cite{Andersen:2002ey}
\ba\la{qhh}
\mathcal{F}_{3qg+4qg}^{(hh)}&=&\frac{5}{144}T^4-\frac{1}{144\pi^2}\bigg[ \frac{1}{\epsilon}+1.2963  \bigg]\bigg( \frac{\mu}{4\pi T}   \bigg)^{4\epsilon} m_D^2T^2\nonumber \\
&&+\frac{1}{16\pi^2}\bigg[ \frac{1}{\epsilon}+8.96751  \bigg] \bigg( \frac{\mu}{4\pi T}   \bigg)^{4\epsilon}m_q^2T^2 \,.
\ea

The $(hh)$ contribution from (\ref{f4s}) can be expanded as
\ba\la{f4shh}
\mathcal{F}_{4s+3gs+4gs}^{(hh)}&=&\sumint_{PQ}\frac{3(d+2)}{P^2Q^2}+M_D^2\sumint_{PQ}\bigg[-\frac{3(5+d)}{P^2Q^4}+\frac{6}{P^2Q^2R^2}    \bigg]  \nonumber \\
 &&+\frac{m_D^2}{d-1}\sumint_{PQ}\bigg\{\frac{3(1-d)}{P^4Q^2} +\frac{3(d-2)}{p^2P^2Q^2}+\frac{3(2+d)}{2p^2Q^2R^2}-\frac{3}{P^2Q^2R^2}\nonumber \\
 &&-\frac{6dq^2}{p^4Q^2R^2} +\frac{6q^2}{p^2P^2Q^2R^2}+\frac{3d(Q\cdot R)}{p^4Q^2R^2}\nonumber \\
 &&+\bigg[\frac{3(1-d)}{p^2P^2Q^2} -\frac{3(1+d)}{2p^2Q^2R^2}+\frac{6dq^2}{p^4Q^2R^2}-\frac{3d(Q\cdot R)}{p^4Q^2R^2}    \bigg] {\mathcal T}_P \bigg\}  \,,
\ea
where $P+R-Q=0$.  Using the sum-integral formulas in Appendix C of Ref.~\cite{Andersen:2003zk} this reduces to
\ba\la{f4shh1}
\mathcal{F}_{4s+3gs+4gs}^{(hh)}&=&\frac{5}{48}T^4-\frac{1}{8\pi^2}\bigg( \frac{1}{\epsilon}+2\gamma+5.72011  \bigg) \bigg( \frac{\mu}{4\pi T}   \bigg)^{4\epsilon}M_D^2T^2\nonumber \\
&&-\frac{7}{384\pi^2} \bigg( \frac{1}{\epsilon}+5.61263  \bigg)  \bigg( \frac{\mu}{4\pi T}   \bigg)^{4\epsilon}m_D^2T^2\,.
\ea
The $(hh)$ contribution to (\ref{fqs}) can be expanded as
\ba\la{fqshh}
\mathcal{F}_{3qs}^{(hh)}&=&24\bigg[\sumint_{P\{Q\}}\frac{-1}{P^2Q^2}+\sumint_{\{PQ\}}\frac{1}{2P^2Q^2}\bigg]+24M_D^2\sumint_{P\{Q\}}\bigg[ \frac{1}{P^4Q^2} -\frac{1}{2P^2Q^2R^2}   \bigg]  \nonumber \\
&&+24m_q^2 \sumint_{P\{Q\}}\bigg\{  \frac{2}{P^2Q^4}+\frac{1}{P^2Q^2R^2} +\frac{p^2-r^2}{P^2Q^2R^2q^2}  +\bigg[ \frac{-1}{P^2Q^2Q_0^2}\nonumber \\
&&+\frac{r^2-p^2}{P^2R^2Q_0^2q^2}    \bigg] {\mathcal T}_Q \bigg\}+ 24m_q^2\sumint_{\{PQ\}}\bigg[\frac{-2}{P^2Q^4}+\frac{1}{P^2Q^2Q_0^2}{\mathcal T}_Q   \bigg]\,,
\ea
where $P+R-Q=0$. Using the sum-integral formulas in Appendix C of Ref.~\cite{Andersen:2003zk} this reduces to
\ba\la{fqshh1}
\mathcal{F}_{3qs}^{(hh)}&=&\frac{5}{48}T^4-\frac{1}{16\pi^2} \bigg( \frac{1}{\epsilon}+5.73824  \bigg)\bigg( \frac{\mu}{4\pi T}   \bigg)^{4\epsilon}M_D^2T^2  \nonumber\\
&&+\frac{3}{16\pi^2}\bigg(  \frac{1}{\epsilon}+9.96751  \bigg)\bigg( \frac{\mu}{4\pi T}   \bigg)^{4\epsilon}m_q^2T^2 \,.
\ea

\subsubsection{ Contributions from the $(hs)$ region}

In the $(hs)$ region, the soft momentum can be any bosonic momentum. The functions that multiply the soft propagators $\Delta_T(0,\textbf{p})$, $\Delta_X(0,\textbf{p})$, or $\Delta_s(0,\textbf{p})$ can be expanded in powers of the soft momentum $\textbf{p}$. In terms involving $\Delta_T(0,\textbf{p})$, the resulting integrals over $\textbf{p}$ have no scale and vanish in dimensional regularization. The integration measure $\int_{\textbf{p}}$ scales like $m_D^3$ for gluon momentum and $M_D^3$ for scalar momentum, respectively.  The soft propagators $\Delta_X(0,\textbf{p})$ and $\Delta_s(0,\textbf{p})$ scale like $1/m_D^2$ and $1/M_D^2$, respectively, and every power of $p$ in the numerator scales like $m_D$ for gluon momentum and $M_D$ for scalar momentum.

The $(hs)$ contribution from the gluonic free energy graphs (\ref{f3g}) is \cite{Andersen:2002ey}
\ba\la{ghs}
\mathcal{F}_{3g+4g+gh}^{(hs)}= -\frac{1}{24\pi}m_D T^3-\frac{11}{384\pi^3}\bigg(\frac{1}{\epsilon}+2\gamma+\frac{27}{11}    \bigg)\bigg( \frac{\mu}{4\pi T}   \bigg)^{2\epsilon}\bigg( \frac{\mu}{2m_D}  \bigg)^{2\epsilon}m_D^3T  \,.
\ea

The $(hs)$ contribution from quark self energy (\ref{fgq}) is
\ba\la{qhs}
\mathcal{F}_{3qg+4qg}^{(hs)}&=&-\frac{1}{12\pi}m_DT^3 + \frac{1}{4\pi^3}m_D m_q^2T\nonumber \\
&&+\frac{1}{48\pi^3}\bigg(\frac{1}{\epsilon}+2\gamma+1+4\log2  \bigg)\bigg( \frac{\mu}{4\pi T}   \bigg)^{2\epsilon}\bigg( \frac{\mu}{2m_D}  \bigg)^{2\epsilon}m_D^3 T\,.
\ea
Note that the sign on the second term differs from Ref.~\cite{Andersen:2003zk}.  This is due to an incorrect sign in the HTL-corrected quark-gluon three vertex in Ref.~\cite{Andersen:2003zk}, which we discuss in the Appendix \ref{subsec:qgv}.

For the $(hs)$ contributions to (\ref{f4s}) and (\ref{fqs}), like QCD, after expansion there will be terms of contributing at ${\cal O}(g)$ and higher. For terms that are already of order $g^3$, we can set $R=Q$ for soft momentum $P$.  For terms that are ${\cal O}(g)$, we must expand the sum-integral to second order in $\textbf{p}$, and then perform the angular integration for $\textbf{p}$, where the linear terms in $\textbf{p}$ vanish and quadratic terms of the form $p^i p^j$ can be replaced by $p^2\delta^{ij}/d$. Therefore, the $(hs)$ contribution from (\ref{f4s}) can be written as
\ba\la{f4shs}
\mathcal{F}_{4s+3gs+4gs}^{(hs)}&=&T\int_{\textbf{p}}\frac{1}{p^2+M_D^2}\sumint_{Q}\bigg[ \frac{3(d+5)}{Q^2}-M_D^2\bigg( \frac{3}{Q^4}+\frac{12q^2}{dQ^6}    \bigg) -m_D^2\frac{3}{Q^4}  \bigg] \nonumber \\
&&+ T\int_{\textbf{p}}\frac{1}{p^2+m_D^2}\sumint_{Q} \bigg\{\frac{6q^2}{Q^4}-\frac{3}{Q^2}+M_D^2\bigg(\frac{9}{Q^4}-\frac{12q^2}{Q^6} \bigg)   \bigg\} \nonumber \\
&&+ m_D^2\bigg[ -\frac{6}{Q^4}+6(d+4)\frac{q^2}{dQ^6} -\frac{24q^4}{dQ^8}\bigg]  ,
\ea
where $P+R-Q=0$.  Using the sum-integral formulas from Appendix C of Ref.~\cite{Andersen:2003zk} this reduces to
\ba\la{f4shs1}
\mathcal{F}_{4s+3gs+4gs}^{(hs)}&=& -T^3\bigg(\frac{m_D}{8\pi}+ \frac{M_D}{2\pi}  \bigg)-\frac{3}{32\pi^3}M_D^2 m_DT\nonumber \\
&&+\bigg[\frac{3}{64\pi^3}M_D m_D^2T+\frac{3}{32\pi^3}M_D^3T\bigg]\bigg( \frac{1}{\epsilon}+2+2\gamma \bigg)\bigg( \frac{\mu}{4\pi T}   \bigg)^{2\epsilon}\bigg( \frac{\mu}{2M_D}  \bigg)^{2\epsilon} \nonumber \\
&&+\frac{1}{128\pi^3} \bigg(\frac{1}{\epsilon}+4+2\gamma \bigg)\bigg( \frac{\mu}{4\pi T}   \bigg)^{2\epsilon}\bigg( \frac{\mu}{2m_D}  \bigg)^{2\epsilon}m_D^3 T\,.
\ea
The $(hs)$ contribution from (\ref{fqs}) can be written as
\ba\la{fqshs}
\mathcal{F}_{3qs}^{(hs)}&=& 24T\int_{\textbf{p}}\frac{1}{p^2+M_D^2}\sumint_{\{Q\}}\bigg[ -\frac{1}{Q^2}+M_D^2\bigg(-\frac{1}{Q^4}+\frac{2q^2}{dQ^6}  \bigg)+m_q^2\frac{2}{Q^4}   \bigg] ,
\ea
where $P+R-Q=0$.  Again using the sum-integral formulas from Appendix C of Ref.~\cite{Andersen:2003zk} this reduces to
\ba\la{fqshs1}
\mathcal{F}_{3qs}^{(hs)}&=& -\frac{1}{4\pi}M_DT^3+\bigg[\frac{3}{16\pi^3} M_D^3T-\frac{3}{4\pi^3}M_D m_q^2T   \bigg]\nonumber \\
&&\hspace{1cm}\times\bigg( \frac{1}{\epsilon}+2+2\gamma+4\log2  \bigg)\bigg( \frac{\mu}{4\pi T}   \bigg)^{2\epsilon}\bigg( \frac{\mu}{2M_D}  \bigg)^{2\epsilon}\,.
\ea

\subsubsection{ Contributions from the $(ss)$ region}

In the $(ss)$ region, all bosonic momentum are soft, and the gluonic HTL correction functions $\mathcal{T}_P$, $\mathcal{T}^{000}$, and $\mathcal{T}^{0000}$ vanish. The gluonic self-energy functions at zero-frequency are $\Pi_T(0,\textbf{p})=0$ and $\Pi_L(0,\textbf{p})=m_D^2$. The scales in the integrals come from the gluonic longitudinal propagator $\Delta_L(0,\textbf{p})=1/(p^2+m_D^2)$ and scalar propagator $\Delta_s(0,\textbf{p})=-1/(p^2+M_D^2)$. Therefore for bosons, in dimensional regularization, at least one such propagator is required in order for the integral to be nonzero, and there is no $(ss)$ contributions coming from fermionic diagrams.

The $(hs)$ contribution to the gluonic free energy graphs (\ref{f3g}) is
\ba\la{gss}
\mathcal{F}_{3g+4g+gh}^{(ss)}= \frac{1}{64\pi^2}\bigg(\frac{1}{\epsilon}+3 \bigg)\bigg( \frac{\mu}{2m_D}  \bigg)^{4\epsilon}m_D^2T^2 \,.
\ea

The $(ss)$ contribution to (\ref{f4s}) can be expanded as
\ba\la{f4sss}
\mathcal{F}_{4s+3gs+4gs}^{(ss)}=T^2\int_{\textbf{pq}}\bigg[\frac{3}{(p^2+m_D^2)(q^2+M_D^2)}+\frac{6M_D^2+9p^2}{p^2(q^2+M_D^2)(r^2+M_D^2)}   \bigg] \,,
\ea
where $P+R-Q=0$. Again using the sum-integral formulas from Appendix C of Ref.~\cite{Andersen:2003zk} this reduces to
\ba\la{f4sss1}
\mathcal{F}_{4s+3gs+4gs}^{(ss)}=\frac{3}{16\pi^2} M_D m_D T^2+\frac{3}{32\pi^2}\bigg( \frac{1}{\epsilon} +8 \bigg)\bigg( \frac{\mu}{2M_D}  \bigg)^{4\epsilon}M_D^2T^2\,. \ea

\section{HTL thermodynamic potential}
\la{combine}

In this section, we calculate the thermodynamic potential $\Omega(T, \lambda, m_D, M_D, m_q, \delta=1)$ explicity, first to LO in the $\delta$ expansion and then to NLO.

\subsection{Leading order}\la{leading}

As we mentioned in Sec.~\ref{potential}, the leading order thermodynamic potential is the sum of the contributions from one-loop diagrams and the leading order vacuum energy counterterm. The contributions come from the one-loop diagrams is the sum of (\ref{gh}), (\ref{qh}), (\ref{sh}), (\ref{gs}) and (\ref{ss}). After multiplying by the appropriate coefficients in (\ref{onel}), one obtains
\ba\la{1op}
\Omega_{\textrm{1-loop}}&=&  \mathcal{F}_{\textrm{ideal}}\bigg\{1-\hat{m}_D^2+4\hat{m}_D^3-6\hat{M}_D^2+24\hat{M}_D^3-8\hat{m_q}^2+16\hat{m_q}^4(6-\pi^2) \nonumber \\
&&\hspace{1.5cm}+\frac{3}{4}\hat{m}_D^4\bigg[\frac{1}{\epsilon}-7+2\gamma+\frac{2\pi^2}{3}+2\log\frac{\hat{\mu}}{2}    \bigg]  +9\hat{M}_D^4\bigg[\frac{1}{\epsilon}+ 2\gamma+2\log\frac{\hat{\mu}}{2} \bigg]   \bigg\} \, , \hspace{1cm}
\ea
where $\mathcal{F}_{\textrm{ideal}}$ is the free energy density of $\mathcal{N}=4$ SYM in the ideal gas limit and $\hat{m}_D$, $\hat{M}_D$, $\hat{m_q}$ and $\hat{\mu}$ are dimensionless variables, defined as
\ba\la{dv}
 \hat{m}_D&=&\frac{m_D}{2\pi T} \, ,  \nonumber \\
 \hat{M}_D&=&\frac{M_D}{2\pi T} \, , \nonumber \\
 \hat{m}_q&=&\frac{m_q}{2\pi T} \, ,   \nonumber \\
 \hat{\mu}&=&\frac{\mu}{2\pi T}  \, .
\ea
Since the leading order vacuum energy counterterm $\Delta_0 \mathcal{E}_0$ should cancel the divergences in the one-loop free energy, we obtain
\ba\la{lct}
\Delta_0 \mathcal{E}_0= d_A \bigg( \frac{1}{128 \pi^2 \epsilon} m_D^4 +\frac{3 }{32\pi^2 \epsilon}M_D^4   \bigg) =\mathcal{F}_{\textrm{ideal}}\bigg(-\frac{3}{4\epsilon} \hat{m}_D^4-\frac{9}{\epsilon}\hat{M}_D^4  \bigg) \,.
\ea
After adding the leading order vacuum renormalization counterterm, our final result for the renormalized LO HTLpt thermodynamic potential  is
\ba\la{1oop}
\Omega_{\textrm{LO}}&=&  \mathcal{F}_{\textrm{ideal}}\bigg\{1-\hat{m}_D^2+4\hat{m}_D^3-6\hat{M}_D^2+24\hat{M}_D^3-8\hat{m}_q^2+16\hat{m}_q^4(6-\pi^2) \nonumber \\
&&\hspace{1.5cm}+\frac{3}{4}\hat{m}_D^4\bigg[-7+2\gamma+\frac{2\pi^2}{3}+2\log\frac{\hat{\mu}}{2}    \bigg]  +18\hat{M}_D^4\bigg[ \gamma+\log\frac{\hat{\mu}}{2} \bigg]   \bigg\}       \,.
\ea

\subsection{Next-to-leading order}

The next-to-leading order corrections to the thermodynamic potential include all of the two-loop free energy diagrams, the gluon, quark, scalar counterterms in Fig.~\ref{2-o}, and the renormalization counterterms. The contributions from the two-loop diagrams include all terms through order $g^5$ is the sum of (\ref{ghh}), (\ref{qhh}), (\ref{f4shh1}), (\ref{fqshh1}), (\ref{ghs}), (\ref{qhs}), (\ref{f4shs1}), (\ref{fqshs1}), (\ref{gss}) and (\ref{f4sss1}) multiplied by $\lambda d_A$. Adding these gives
\ba\la{2op}
\Omega_{\textrm{2-loop}}&=& \mathcal{F}_{\textrm{ideal}} \frac{\lambda}{\pi^2} \bigg\{-\frac{3}{2}+3\hat{m}_D+9\hat{M}_D-\frac{9}{2}\hat{M}_D\hat{m}_D+\frac{9}{2}\hat{m}_D \hat{M}_D^2  \nonumber \\
&&-12\hat{m}_D \hat{m}_q^2 +\frac{3}{8} \hat{m}_D^2 \bigg[\frac{1}{\epsilon}+ 4\log\hat{m}_D+4\log\frac{\hat{\mu}}{2} + 5.93198468  \bigg] \nonumber \\
&&+\frac{9}{4} \hat{M}_D^2\bigg[ \frac{1}{\epsilon}+ 4\log\hat{M}_D+4\log\frac{\hat{\mu}}{2}  +4.99154798         \bigg]+\frac{1}{8}\hat{m}_D^3\bigg[7-32\log2   \bigg]\nonumber \\
&&-6\hat{m}_q^2 \bigg[\frac{1}{\epsilon}+4\log\frac{\hat{\mu}}{2}  +9.71751112   \bigg]-36\log2 \hat{M}_D^3+144\log2\hat{M}_D\hat{m}_q^2 \nonumber \\
&&+\bigg[-\frac{27}{2} \hat{M}_D^3-\frac{9}{4} \hat{M}_D\hat{m}_D^2+36\hat{M}_D\hat{m}_q^2\bigg]\bigg[2+\frac{1}{\epsilon}+2\gamma-2\log\hat{M}_D+4\log\frac{\hat{\mu}}{2}\bigg]
    \bigg\} . \hspace{7mm}
\ea

The total NLO HTL counterterm contribution is the sum of (\ref{gcth}), (\ref{qcth}), (\ref{scth}), (\ref{gcts}) and (\ref{scts}) multiplied by the Casimirs in (\ref{nelx1})
\ba\la{cou2}
\Omega_{\textrm{HTL}}&=& \mathcal{F}_{\textrm{ideal}} \bigg\{\hat{m}_D^2-6\hat{m}_D^3+6\hat{M}_D^2-36\hat{M}_D^3+8\hat{m}_q^2+32\hat{m}_q^4(\pi^2-6)  \nonumber \\
&&\hspace{5mm}-\frac{3}{2}\hat{m}_D^4\bigg[\frac{1}{\epsilon}-7+2\gamma+2\log\frac{\hat{\mu}}{2}+\frac{2\pi^2}{3}  \bigg]-18\hat{M}_D^4\bigg[\frac{1}{\epsilon}+2\gamma+2\log\frac{\hat{\mu}}{2}\bigg] \bigg\} . \hspace{7mm}
\ea
The ultraviolet divergences in (\ref{2op}) and (\ref{cou2}) will be removed by the renormalization of the vacuum energy density $\mathcal{E}_0$ and the HTL mass parameters $m_D$, $M_D$, and $m_q$. The renormalization counterterm contribution at linear order in $\delta$ is denoted by $\Delta\Omega_{\rm NLO}$ in Eqs.~\eqref{nel} and (\ref{nelx2}). We cannot obtain its form directly from Ref.~\cite{Andersen:2003zk} due to the fact that there are contributions coming from the scalar fields in $\mathcal{N}=4$ SYM theory, but we can use the same method as in QCD. The form of $\Delta_1 \mathcal{E}_0$ can be obtained by expanding (\ref{dc0}) to first order in $\delta$, which is
\ba\la{dc01}
\Delta_1 \mathcal{E}_0=-d_A\bigg(\frac{m_D^4}{64\pi^2 \epsilon}+\frac{3M_D^4}{16\pi^2 \epsilon}\bigg) =\mathcal{F}_{\textrm{ideal}} \bigg( \frac{3}{2\epsilon}\hat{m}_D^4+\frac{18}{\epsilon}\hat{M}_D^4  \bigg) .
\ea
This renormalization counterterm cancels the divergences in $\Omega_{\textrm{HTL}}$ \eqref{cou2}. For the two-loop self energy, as we can see, the divergent terms are
\ba
\mathcal{F}_{\textrm{ideal}} \frac{\lambda}{\pi^2} \bigg[ \frac{3}{8\epsilon} \hat{m}_D^2 +\frac{9}{4\epsilon} \hat{M}_D^2 - \frac{6}{\epsilon}\hat{m}_q^2 -\frac{27}{2\epsilon} \hat{M}_D^3-\frac{9}{4\epsilon} \hat{M}_D\hat{m}_D^2+ \frac{36}{\epsilon}\hat{M}_D\hat{m}_q^2 \bigg].
\ea
Since there are two mixed terms $\hat{M}_D\hat{m}_D^2$ and $\hat{M}_D\hat{m}_q^2$ which is a big difference from QCD, we cannot use the following formula
\ba
\Delta_1m_D^2\frac{\partial}{\partial m_D^2} \Omega_{\textrm{LO}} +\Delta_1m_q^2\frac{\partial}{\partial m_q^2} \Omega_{\textrm{LO}}+\Delta_1M_D^2\frac{\partial}{\partial M_D^2} \Omega_{\textrm{LO}}.
\ea
This is because we cannot get the two mixed term. In order to cancel the ultraviolet divergence for two-loop self energy, the simplest form is in Eq.~(\ref{nelx2}).  Using (\ref{dmass}), (\ref{dc01}), and (\ref{nelx2}), one finds
\ba\la{dgo1}
\Delta\Omega_{\rm NLO}&=& \mathcal{F}_{\textrm{ideal}}\bigg\{\frac{3}{2\epsilon} \hat{m}_D^4 + \frac{18}{\epsilon} \hat{M}_D^4+ \frac{\lambda}{\pi^2}\bigg[\bigg(-\frac{3}{8}\hat{m}_D^2 -\frac{9}{4} \hat{M}_D^2+6\hat{m}_q^2 \bigg) \nonumber \\
&& \hspace{1cm} \times  \bigg(\frac{1}{\epsilon}+2+2\frac{\zeta'(-1)}{\zeta(-1)}+2\log\frac{\hat{\mu}}{2}   \bigg)  +\bigg(\frac{27}{2} \hat{M}_D^3+\frac{9}{4}\hat{M}_D\hat{m}_D^2-36\hat{M}_D\hat{m}_q^2 \bigg) \nonumber \\
&& \hspace{1cm} \times \bigg(\frac{1}{\epsilon}+ 2-2\log\hat{M}_D+2\log\frac{\hat{\mu}}{2}   \bigg)   \bigg] \bigg\} \,.
\ea

Adding the leading order thermodynamic potential in (\ref{1oop}), the two-loop free energy in (\ref{2op}), the HTL gluon and quark counterterms in (\ref{cou2}), and the HTL vacuum and mass renormalizations in (\ref{dgo1}), our final expression for the NLO HTLpt thermodynamic potential in $\mathcal{N}=4$ SYM is
\ba\la{loo2p}
\Omega_{\textrm{NLO}}&=& \mathcal{F}_{\textrm{ideal}} \bigg\{ 1 -2\hat{m}_D^3-12\hat{M}_D^3+16\hat{m}_q^4(\pi^2-6)-18\hat{M}_D^4\bigg(\gamma+\log\frac{\hat{\mu}}{2} \bigg) \nonumber \\
&&  -\frac{3}{2}\hat{m}_D^4\bigg( -\frac{7}{2}+\gamma+\frac{\pi^2}{3}+\log\frac{\hat{\mu}}{2}  \bigg) +\frac{\lambda}{\pi^2}\bigg[-\frac{3}{2}+3\hat{m}_D+9\hat{M}_D\nonumber \\
&& -\frac{9}{2}\hat{m}_D\hat{M}_D+\frac{9}{2}\hat{m}_D \hat{M}_D^2-12\hat{m}_D \hat{m}_q^2 -12\hat{m}_q^2 \bigg(1.87370184+\log\frac{\hat{\mu}}{2}     \bigg)\nonumber \\
&& +\frac{3}{4} \hat{m}_D^2 \bigg(-0.01906138 + 2\log\hat{m}_D+\log\frac{\hat{\mu}}{2}   \bigg) +\frac{1}{8}\hat{m}_D^3\bigg(7-32\log2   \bigg)\nonumber \\
&& +\frac{9}{2}\hat{M}_D^2\bigg( -0.489279733 + 2\log\hat{M}_D+\log\frac{\hat{\mu}}{2}     \bigg)-\frac{9}{2} \hat{M}_D\hat{m}_D^2\bigg(\gamma+\log\frac{\hat{\mu}}{2}\bigg)\nonumber \\
&&   + 72\hat{M}_D\hat{m}_q^2\bigg(\gamma+2\log2+\log\frac{\hat{\mu}}{2} \bigg)- 9\hat{M}_D^3 \bigg(3\gamma+4\log2+3\log\frac{\hat{\mu}}{2}  \bigg)  \bigg]         \bigg\} \,.
\ea
Note that this result reproduces the perturbative expansion given in \eqref{weaexp} through ${\cal O}(\lambda^{3/2})$ in the weak-coupling limit.  This can be verified by taking $\hat{m}_D$, $\hat{M}_D$, and $\hat{m}_q$ to be given by their leading-order expressions \eqref{eq:pertmasses} and truncating the resulting expansion in the \mbox{'t Hooft} coupling at ${\cal O}(\lambda^{3/2})$.

\subsection{Gap equations}

The gluon, scalar, and quark mass parameters $m_D$, $M_D$, and $m_q$ are determined by using the variational method, requiring that the derivative of $\Omega_{\textrm{NLO}}$ with respect to each parameter is zero
\ba\la{gap1}
\frac{\partial}{\partial m_q}\Omega_{\textrm{NLO}}(T, \lambda, m_D, M_D, m_q, \delta=1)&=&0 \, ,    \nonumber \\
\frac{\partial}{\partial m_D}\Omega_{\textrm{NLO}}(T, \lambda, m_D, M_D, m_q, \delta=1)&=&0 \, ,    \nonumber \\
\frac{\partial}{\partial M_D}\Omega_{\textrm{NLO}}(T, \lambda, m_D, M_D, m_q, \delta=1)&=&0    \,.
\ea

\begin{figure}[t]
\centering
\includegraphics[width=0.475\textwidth]{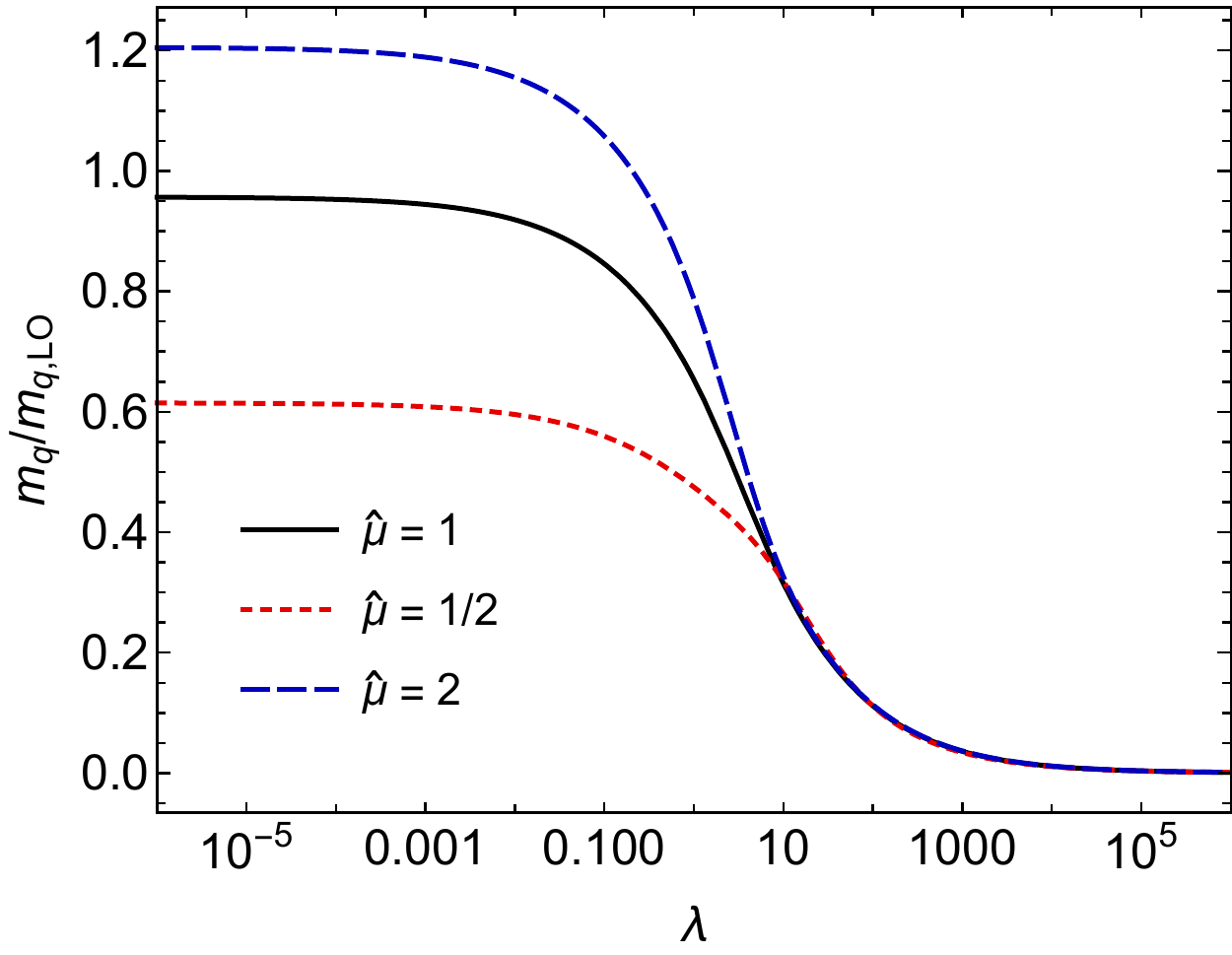}\;\;\;
\includegraphics[width=0.475\textwidth]{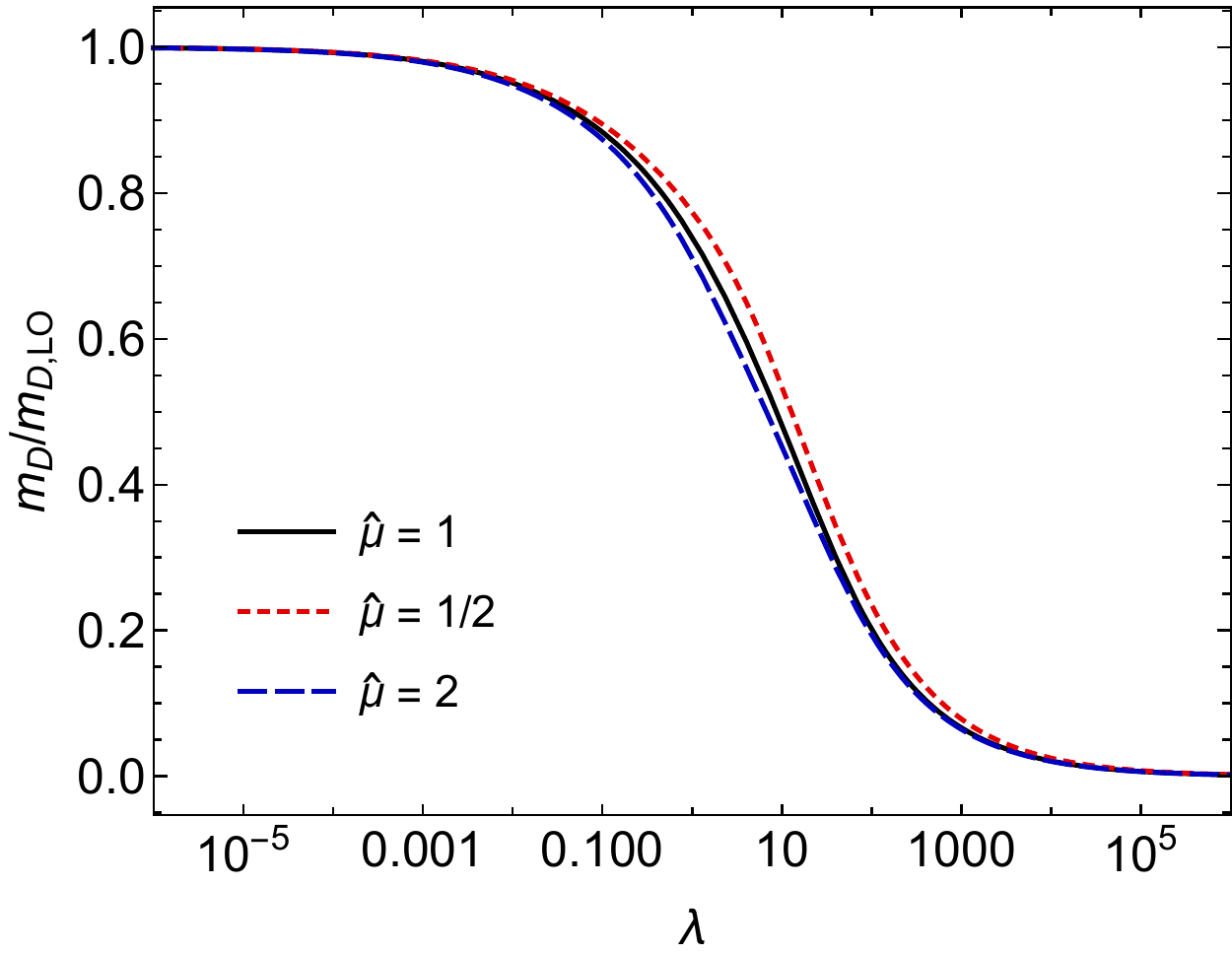}\\[2ex]
\includegraphics[width=0.475\textwidth]{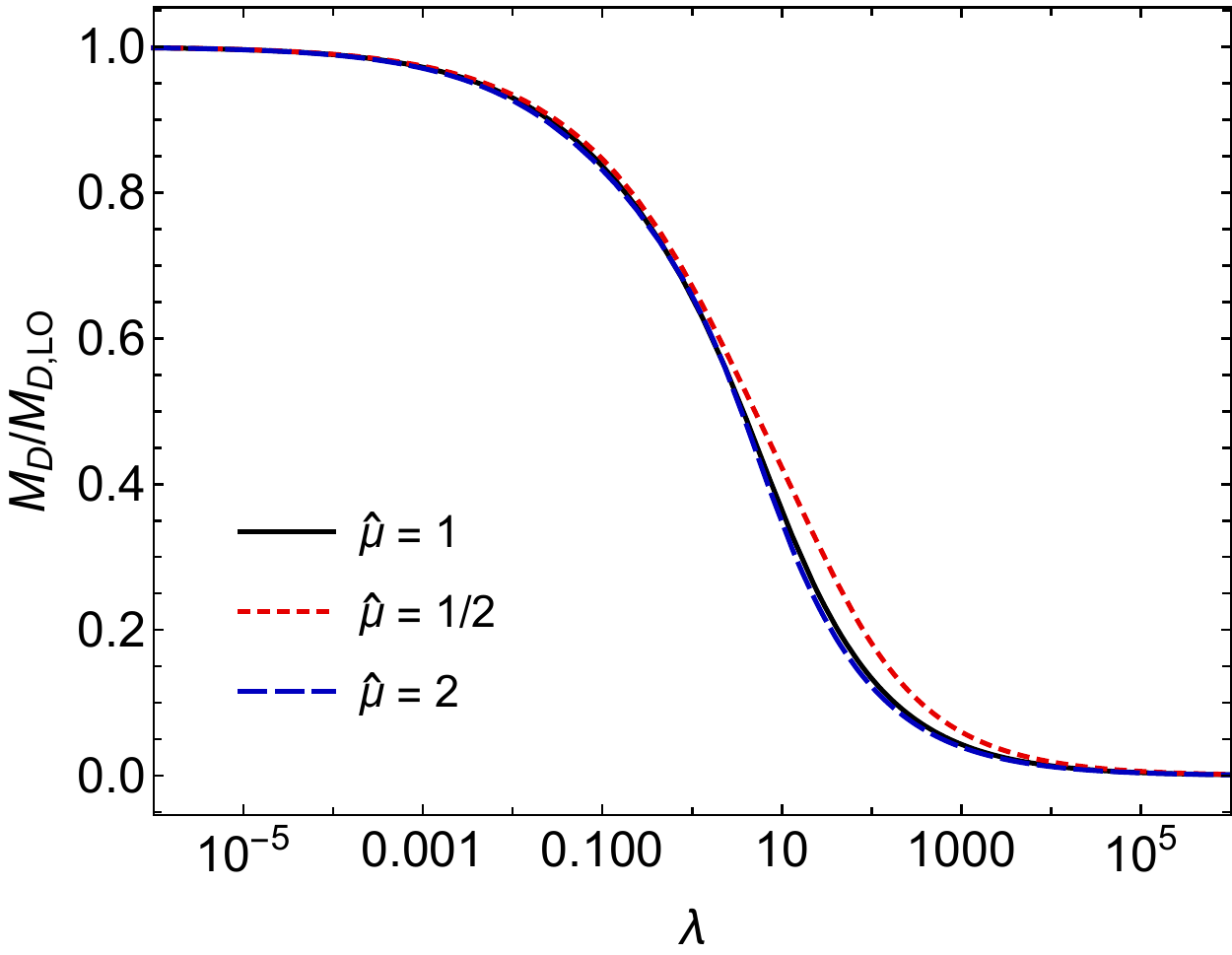}
\vspace*{-.08cm}
\caption{\label{gapsol}
Numerical solution of gap equations for $m_q$, $m_D$, and $M_D$ as a function of $\lambda$. In each panel the results are scaled by their corresponding leading-order weak-coupling limits.}
\end{figure}

\noindent
The first equation gives
\be\la{gapmq}
\hat{m}_q^2\big(\pi^2-6 \big)=\frac{\lambda}{4\pi^2}\bigg[\frac{3}{2}\hat{m}_D+\frac{3}{2} \bigg(1.87370184+\log\frac{\hat{\mu}}{2}   \bigg) - 9 \hat{M}_D\bigg( \gamma+2\log2+\log\frac{\hat{\mu}}{2} \bigg) \bigg] .
\ee
The second equation gives
\ba\la{gapmd}
&&\hspace{-2mm}\hat{m}_D^2+\hat{m}_D^3\bigg(\!{-}\frac{7}{2}+\gamma+\frac{\pi^2}{3}+ \log\frac{\hat{\mu}}{2} \bigg) \!=\! \frac{\lambda}{4\pi^2}\bigg[2-3\hat{M}_D +3\hat{M}_D^2 -8\hat{m}_q^2  +\frac{7}{4}\hat{m}_D^2\bigg(\!1-\frac{32}{7}\log2   \bigg) \nonumber \\
&&  \hspace{15mm} -6\hat{m}_D\hat{M}_D\bigg(\gamma+ \log\frac{\hat{\mu}}{2}  \bigg) +\hat{m}_D\bigg( 0.980939+2\log\hat{m}_D+ \log\frac{\hat{\mu}}{2}   \bigg)\bigg] .
\ea
The third equation gives
\ba\la{gapm}
&&\hat{M}_D^2+2\hat{M}_D^3\bigg( \gamma+ \log\frac{\hat{\mu}}{2} \bigg)= \frac{\lambda}{4\pi^2}\bigg[1-\frac{1}{2}\hat{m}_D +\hat{m}_D\hat{M}_D-\frac{1}{2}\hat{m}_D^2\bigg(\gamma+ \log\frac{\hat{\mu}}{2}  \bigg)\nonumber \\
&& \hspace{2cm} - 3\hat{M}_D^2\bigg(4\log2+3\gamma +3\log\frac{\hat{\mu}}{2}\bigg)+ 8\hat{m}_q^2\bigg(\gamma+ 2\log2+ \log\frac{\hat{\mu}}{2}  \bigg) \nonumber \\
&& \hspace{2.5cm} + \hat{M}_D\bigg(0.51072+2 \log\hat{M}_D+\log\frac{\hat{\mu}}{2} \bigg)         \bigg] .
\ea
Note that the terms proportional to $\hat{m}_q^2$ in Eqs.~(\ref{gapmd}) and (\ref{gapm}) can be written in terms of $\hat{M}_D$ and $\hat{m}_D$ by using (\ref{gapmq}).

In practice, one must solve these three equations simultaneously in order to obtain the gap equation solutions for $\hat{m}_q^2(\lambda)$, $\hat{m}_D^2(\lambda)$, and $\hat{M}_D^2(\lambda)$.  In Fig.~\ref{gapsol} we present our numerical solutions to these three gap equations scaled by the corresponding leading-order weak-coupling limits
\ba
\hat{m}_{q,\textrm{pert}}^2 &=& \frac{\lambda}{8\pi^2} \, ,  \nonumber \\
\hat{m}_{D,\textrm{pert}}^2 &=& \frac{\lambda}{2\pi^2} \, ,  \nonumber \\
\hat{M}_{D,\textrm{pert}}^2 &=&\frac{\lambda}{4\pi^2} \, .
\la{eq:pertmasses}
\ea
In all three panels, the black line is the solution when taking the renormalization scale $\hat\mu =1$, the red dashed line is $\hat\mu=1/2$, and the blue long-dashed line is $\hat\mu=2$. As can be seen from Fig.~\ref{gapsol}, the gap equation solution for $\hat{m}_q$ does not approach its perturbative limit when $\lambda$ is approaches zero.  This is similar to what was found in NLO HTLpt applied to QCD \cite{Andersen:2003zk}.

\section{Thermodynamic functions}\la{entropy}

The NLO HTLpt approximation to the free energy is obtained by evaluating the NLO HTLpt thermodynamic potential (\ref{loo2p}) at the solution of the gap equations (\ref{gap1})
\be
\mathcal{F}_{\rm NLO} = \Omega_{\rm NLO}(T, \lambda, m_D^{\rm gap}, M_D^{\rm gap}, m_q^{\rm gap}, \delta=1) \, .
\ee
The pressure, entropy density, and energy density can then be obtained using
\ba
\mathcal{P} &=& -\mathcal{F} \, , \nonumber \\
\mathcal{S} &=& -\frac{d \mathcal{F}}{d T}  \, , \nonumber \\
\mathcal{E} &=& \mathcal{F} - T\frac{d \mathcal{F}}{d T} \, .
\ea
Note that due the conformality of the SYM theory, in all three of these functions, the only dependence on $T$ is contained in the overall factor of $\mathcal{F}_{\rm ideal}$.  As a result, when scaled by their ideal limits, the ratios of all of these quantities are the same, i.e. $\mathcal{P}/\mathcal{P}_{\rm ideal}$ = $\mathcal{S}/\mathcal{S}_{\rm ideal}$ = $\mathcal{E}/\mathcal{E}_{\rm ideal}$.

\begin{figure}[t!]
\centering
\includegraphics[width=0.9\textwidth]{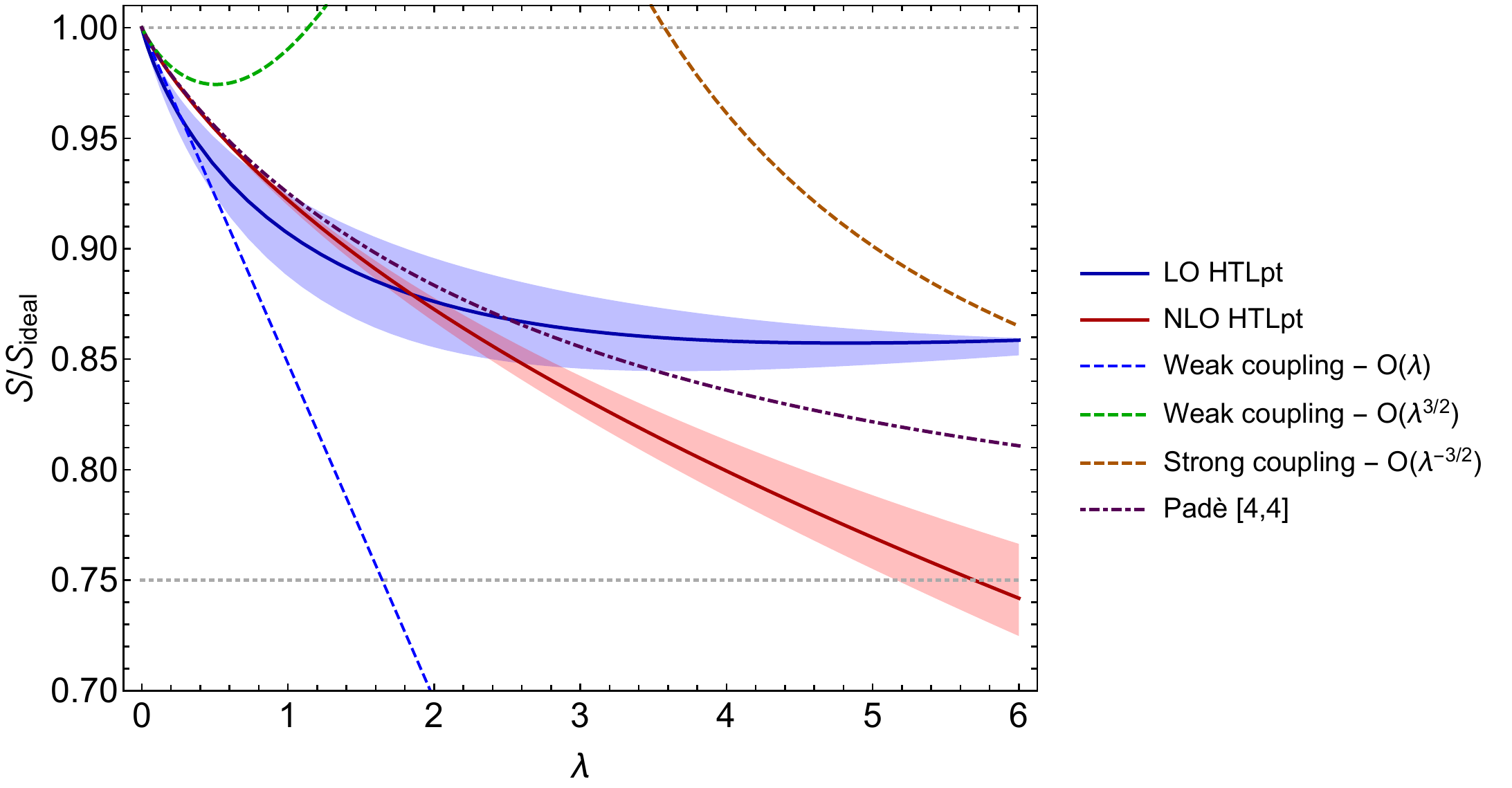}
\vspace*{-.08cm}
\caption{\label{LONLO}
Comparison of the LO and NLO HTLpt results for the scaled entropy density with prior results from the literature.  A detailed description of the various lines can be found in the text.}
\end{figure}

\subsection{Numerical results}\la{results}

In Fig.~\ref{LONLO} we present our final results for the scaled entropy density in $\mathcal{N}=4$ SYM.  The red solid line with a red shaded band is the NLO HTLpt result and the blue solid line with a blue shaded band is the LO HTLpt result determined by evaluating Eq.~\eqref{1oop} at the solution to the NLO mass gap equations \eqref{gap1}.  The HTLpt shaded bands result from variation of the renormalization scale $\hat\mu$.  Herein, we take $\hat\mu \in \{1/2, 1, 2\}$ with the central value of the renormalization scale plotted as solid blue and red lines for the LO and NLO results, respectively.  The blue dotted line is the weak-coupling result \eqref{weaexp} truncated at order $\lambda$, the green dotted line is the weak-coupling result  \eqref{weaexp} truncated at order $\lambda^{3/2}$, the dark-orange dotted line is the strong-coupling result \eqref{stro}.  The purple dot-dashed line is the result of constructing a $R_{[4,4]}$ Pad\'{e} approximant which interpolates between the weak and strong coupling limits \cite{Blaizot:2006tk}.  Finally, the grey dotted lines indicate the strong and weak coupling limits of 3/4 and 1, respectively.

As can be seen from Fig.~\ref{LONLO}, the LO and NLO HTLpt predictions are close to one another out to $\lambda \lesssim 2$.  Computing the ratio of the NLO and LO results, we find that they are within $\sim 5\%$ of one another in this range.  This is a much smaller change from LO to NLO than is found using the naive weak-coupling expansion.  We also observe that the size of the scale variation (shown as shaded red and blue bands) decreases as one goes from LO to NLO.  Comparing the bands at $\lambda =1$ we find that the LO HTLpt variation around $\hat\mu=1$ is on the order of 2\%, whereas the NLO order HTLpt variation is 0.3\%.  For $\lambda \gtrsim 6$ the NLO HTLpt is below the value expected in the strong coupling limit.  At smaller couplings, $\lambda \lesssim 1$ we observe that the NLO HTLpt result is very close to the $R_{[4,4]}$ Pad\'{e} approximant.  This could be coincidental, however, it is suggestive that somehow the $R_{[4,4]}$ Pad\'{e} approximant may provide a reasonable approximation to $\mathcal{N}=4$ SYM thermodynamics despite its ad hoc construction.

\begin{figure}[t!]
\centering
\includegraphics[width=0.9\textwidth]{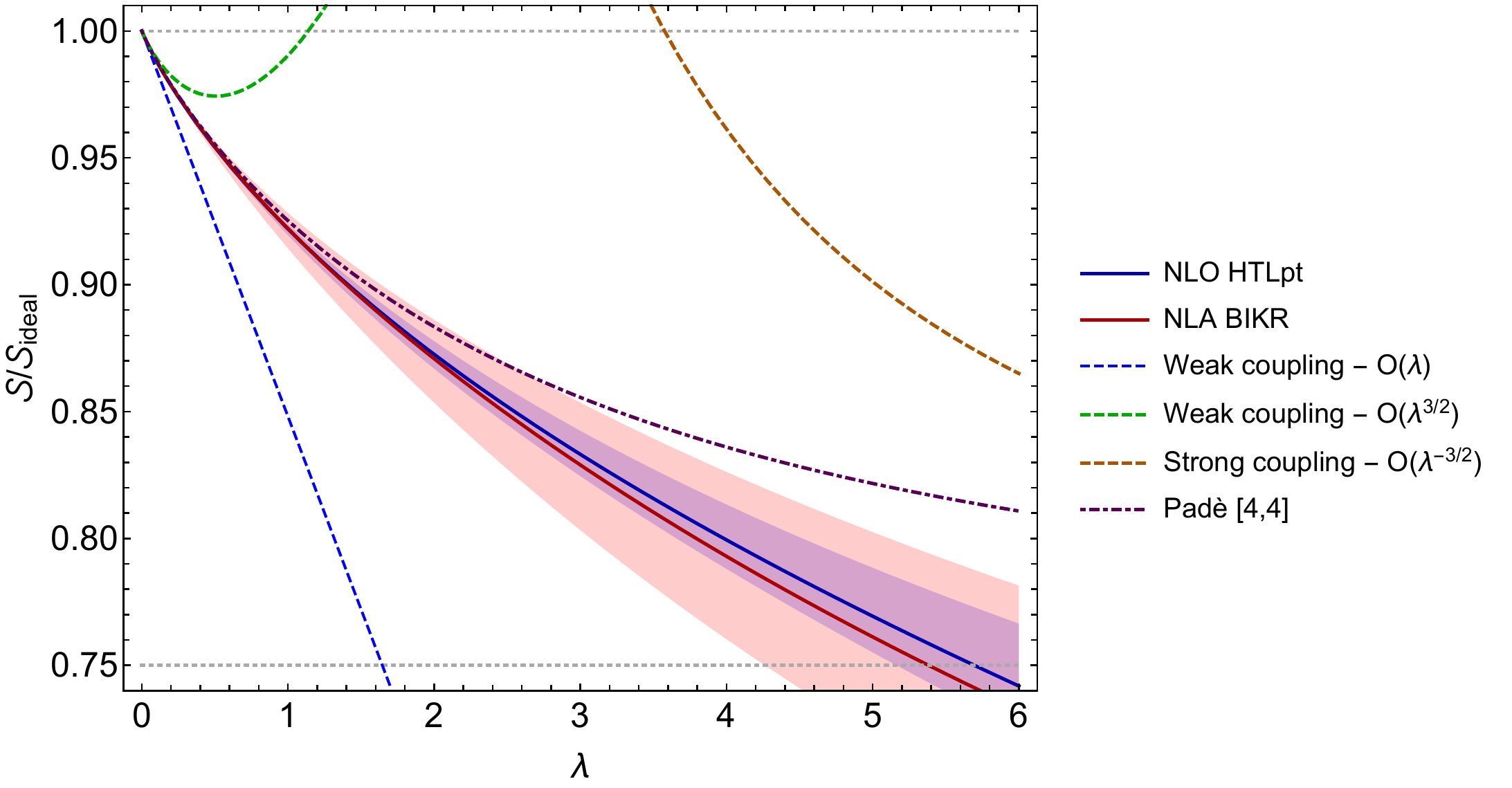}
\vspace*{-.08cm}
\caption{\label{HTLNLA1}
Comparison of our NLO HTLpt result for the scaled entropy density with the prior NLA work of Blaizot, Iancu, Kraemmer, and Rebhan (BIKR) \cite{Blaizot:2006tk}. A detailed description of the various lines can be found in the text. }
\end{figure}

A similar conclusion was obtained in a prior study of HTL resummation in $\mathcal{N}=4$ SYM thermodynamics \cite{Blaizot:2006tk}.  In their work, Blaizot, Iancu, Kraemmer, and Rebhan (BIKR) used the $\Phi$-derivable framework to obtain an approximately self-consistent approximation to the scaled entropy density.  In their approach, the one-loop $\Phi$-derivable result for the entropy density has approximate next-to-leading-order accuracy (NLA), so it should be comparable to the our NLO HTLpt result.  In Fig.~\ref{HTLNLA1} we present a comparison of our NLO HTLpt result with the BIKR NLA result~\cite{Blaizot:2006tk}.  In this Figure, the blue line with a blue shaded band is our NLO HTLpt result and the solid red line with a red shaded band is the NLA result from Ref.~\cite{Blaizot:2006tk}.  The dashed and dotted lines are the same as the previous figure.  We find that, for $\hat{\mu}=1$, the two calculations are within $\leq 2\%$ of one another for $\lambda \lesssim 6$.  We observe that the NLO HTLpt result has a smaller scale variation than the NLA result at all couplings shown.

\begin{figure}[t!]
\centering
\includegraphics[width=0.9\textwidth]{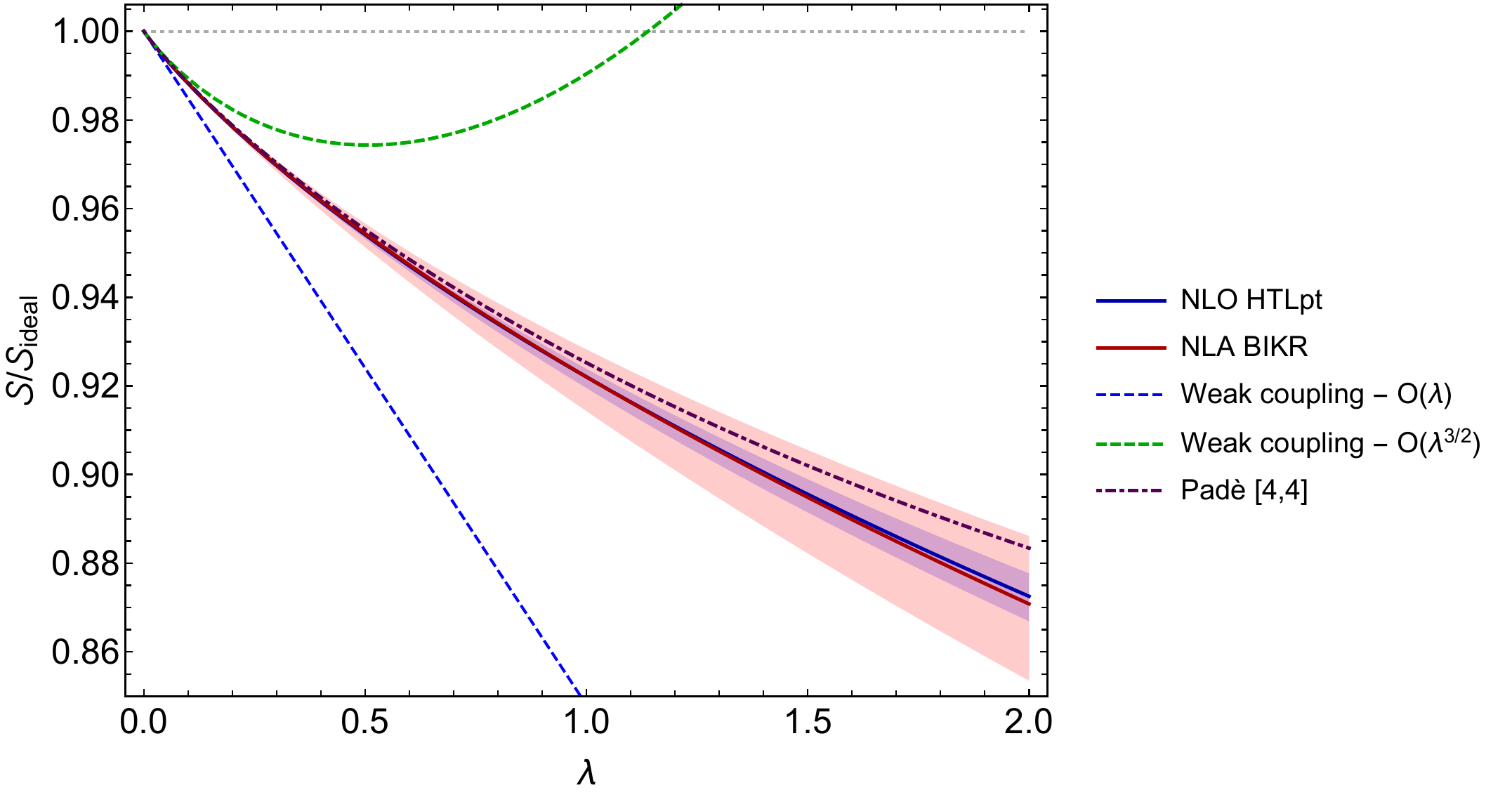}
\vspace*{-.08cm}
\caption{\label{HTLNLA2}
Comparison of our NLO HTLpt result for the scaled entropy density with prior results at small $\lambda$.  Lines are the same as in Fig.~\ref{HTLNLA1}. }
\end{figure}

Finally, in Fig.~\ref{HTLNLA2} we present a comparison of all results for $\lambda \leq 2$.  The various weak-coupling lines in this Figure are the same as in Fig.~\ref{LONLO}.  As can be seen from this Figure, there is excellent agreement between the NLA calculation of BIKR and NLO HTLpt in this coupling range.  We also see that the $R_{[4,4]}$ Pad\'{e} approximant overlaps with both calculations at smaller $\lambda$.  Given the agreement between our NLO results and the BIKR NLA results in this range of 't Hooft coupling, one can try to estimate the range of temperatures this might map to in a real-world QGP.  This is a fraught endeavor, however, since one can choose to match a variety of quantities with, to our knowledge, no unique prescription.  In Ref.~\cite{Blaizot:2006tk} the authors advocated matching the scaled entropy density.  For the purposes of a ball-park estimate, we will follow their suggestion.  State-of-the-art lattice data for the scaled entropy density indicates that corrections to the ideal limit saturate above approximately $T \sim 3 T_c \sim 450$ MeV at value of ${\cal S}_{\rm QCD}/{\cal S}_{\rm QCD,0} \sim 0.8-0.85$ \cite{Borsanyi:2010cj}.  Matching this to the same ratio in $\mathcal{N}=4$ SYM, one finds from Fig.~\ref{HTLNLA1} that this requires $\lambda \sim 3-4$ using $\hat\mu =1$.\footnote{Note that all such estimates should be taken with care since in QCD, unlike $\mathcal{N}=4$ SYM, there is conformal symmetry breaking which causes, e.g.,  ${\cal S}_{\rm QCD}/{\cal S}_{\rm QCD,0} \neq {\cal P}_{\rm QCD}/{\cal P}_{\rm QCD,0}$.   If one were to use the scaled pressure instead, one for find a different limit for $\lambda$.  Additionally, our choice of $\hat\mu =1$ is somewhat arbitrary and varying this scale will result in further variation of the constraint on the effective 't Hooft coupling.}  Our results suggest that the NLO HTLpt result for $\mathcal{N}=4$ SYM can be trusted to with high accuracy for $\lambda \lesssim 2$.  This provides motivation for extending our calculation to NNLO.

\section{Conclusions}

In this paper we have extended the LO and NLO HTLpt calculation of the thermodynamic potential in QCD to $\mathcal{N}=4$ SYM theory. We have presented results for the LO and NLO HTLpt predictions for the thermodynamics of $\mathcal{N}=4$ SYM for arbitrary $N_c$.  We found that it is possible to extend the range of applicability of perturbative calculations of thermodynamics in $\mathcal{N}=4$ SYM theory to intermediate couplings, albeit using involved resummations.  We compared our NLO HTLpt results to approximately self-consistent resummations obtained previously in Ref.~\cite{Blaizot:2006tk} and found them to be in excellent agreement with our NLO HTLpt results for the scaled entropy density for $\lambda \lesssim 6$.  Compared to the method used in Ref.~\cite{Blaizot:2006tk}, our HTLpt results are manifestly gauge-invariant and the HTLpt framework allows for systematic extension of the calculation to higher loop orders.  It would be interesting to extend the HTLpt results obtained here to NNLO as has been done in QCD.  For this purpose, it seems necessary to first establish the naive perturbative corrections to SUSY thermodynamics at orders $g^4$ and $g^5$.  Work along these lines is in progress.

\acknowledgments

We thank A. Rebhan for providing us with the Mathematica notebook used to generate the final results from Ref.~\cite{Blaizot:2006tk}.  Q.D. was supported by the China Scholarship Council under Project No.~201906770021.  M.S. and U.T. were supported by the U.S. Department of Energy, Office of Science, Office of Nuclear Physics under Award No.~DE-SC0013470.

\appendix

\section{HTL Feynman rules for $\mathcal{N}=4$ SYM}\la{fmr}

In this appendix, we will present the Feynman rules for HTLpt applied to $\mathcal{N}=4$ SYM. The Feynman rules are given in Minkowski space to facilitate future applications to real-time processes. A Minkowski momentum is denoted by $p=(p_0,\textbf{p})$, and satisfies $p\cdot q =p_0 q_0-\textbf{p}\cdot \textbf{q}$. The vector that specifies the thermal rest frame is $n=(1,\textbf{0})$.

\subsection{Gluon polarization tensor}\la{gsfn}

In $\mathcal{N}=4$ SYM, there are six diagrams that contribute to the LO gluon self energy. In the HTL limit, for massless bosons and fermions, the gluon polarization tensor was derived in Ref.~\cite{Czajka:2012gq}
\ba\la{sfgy}
\Pi_{ab}^{\mu\nu}(p)=-g^2 N_c \delta_{ab}\int\frac{d^3 k}{(2\pi)^3} \frac{f(\textbf{k})}{|\textbf{k}|}\frac{p^2 k^\mu k^\nu-(p\cdot k)(p^\mu k^\nu+k^\mu p^\nu)+(p\cdot k)^2g^{\mu\nu}}{(p\cdot k)^2}    \, ,
\ea
where $f(\textbf{k})\equiv 2n_g(\textbf{k})+8n_q(\textbf{k})+6n_s(\textbf{k})$ is the effective one-particle distribution function for an \mbox{$\mathcal{N}=4$} SYM theory. The coefficients of $n_g$, $n_q$, $n_s$ are equal to the number of degrees of freedom of the gauge field, fermions, and scalars. Since $\Pi_{ab}^{\mu\nu}(p)$ is symmetric and transverse in its Lorentz indices, it is gauge independent.

We can define the Debye mass for the gauge field using
\ba\la{md}
m_D^2=-g_{\mu\nu}\Pi_{aa}^{\mu\nu}(p)=2g^2 N_c \int\frac{d^3 k}{(2\pi)^3} \frac{f(\textbf{k})}{|\textbf{k}|}=2\lambda T^2, \quad \lambda=g^2N_c   \, ,
\ea
which has been given previously in Refs.~\cite{Kim:1999sg} and \cite{Czajka:2012gq}. Then using integration by parts from Ref.~\cite{Mrowczynski:2004kv} applied to (\ref{sfgy}), we obtain
\ba
\Pi^{\mu\nu}(p)=-g^2 N_c \int\frac{d^3 k}{(2\pi)^3}\frac{\partial f(\textbf{k})}{\partial|\textbf{k}|} \bigg[y^\mu y^\nu\frac{ p \cdot n }{p\cdot y} -n^\mu n^\nu  \bigg]   ,
\ea
where $y^\mu\equiv k^\mu/|\textbf{k}|=(1,\textbf{k}/|\textbf{k}|)\equiv(1,\hat{\textbf{y}})$. After integration over the length of momentum $|\textbf{k}|$, the HTL gluon polarization tensor can be written as
\ba\la{sfg}
\Pi^{\mu\nu}(p)=m_D^2\big[ \mathcal{T}^{\mu\nu}(p,-p)-n^\mu n^\nu   \big] .
\ea
Where we have introduced a rank-two tensor $\mathcal{T}^{\mu\nu}(p,q)$ which is defined only when $p+q=0$ as
\ba\la{ts}
\mathcal{T}^{\mu\nu}(p,-p)= \bigg\langle y^\mu y^\nu \frac{p \cdot n}{p \cdot y} \bigg \rangle_{\hat{\textbf{y}}}      \,.
\ea
The angular brackets indicate averaging over the spatial direction of the light-like vector $y$. The tensor $\mathcal{T}^{\mu\nu}$ is symmetric in $\mu$ and $\nu$, and satisfies the ``Ward identity''
\ba\la{wit}
p_\mu\mathcal{T}^{\mu\nu}(p,-p)=(p\cdot n) n^\nu    \,.
\ea
As a result, the polarization tensor $\Pi^{\mu\nu}$ is also symmetric in  $\mu$ and $\nu$ and satisfies
\ba\la{wis}
p_\mu \Pi^{\mu\nu}(p)&=& 0,   \nonumber \\  g_{\mu\nu}\Pi^{\mu\nu}(p)&=&-m_D^2 \,.
\ea

The gluon polarization tensor can also be expressed in terms of two scalar functions, the transverse and longitudinal polarization functions $\Pi_T$ and $\Pi_L$, defined by
\ba\la{tslo0}
 \Pi_T(p)&=& \frac{1}{d-1}\big( \delta^{ij}-\hat{p}^i\hat{p}^j  \big) \Pi^{ij}(p),  \nonumber \\
 \Pi_L(p)&=& -\Pi^{00}(p)  \, ,
\ea
where $\hat{\textbf{p}}=\textbf{p}/|\textbf{p}| $ is the unit vector in the direction of $\textbf{p}$. The gluon polarization tensor can be written in terms of these two functions
\ba\la{sfg1}
\Pi^{\mu\nu}(p)=- \Pi_T(p)T_p^{\mu\nu} -\frac{1}{n_p^2}\Pi_L(p) L_p^{\mu\nu} \, ,
\ea
where the tensor $T_p$ and $L_p$ are
\ba\la{tl}
 T_p^{\mu\nu}&=& g^{\mu\nu}-\frac{p^\mu p^\nu}{p^2}- \frac{n_p^\mu n_p^\nu}{n_p^2},  \nonumber \\    L_p^{\mu\nu}&=& \frac{n_p^\mu n_p^\nu}{n_p^2} \, ,
\ea
Above, the four-vector $n_p^\mu$ is
 \ba\la{np}
 n_p^\mu=n^\mu- \frac{n\cdot p}{p^2}p^\mu\,,
\ea
which satisfies $p\cdot n_p=0$ and $n_p^2=1-(n\cdot p)^2/p^2$. Then (\ref{wis}) reduces to the identity
\ba\la{wis1}
(d-1)\Pi_T(p)+\frac{1}{n_p^2}\Pi_L(p) =m_D^2   \, ,
\ea
at the same time, we can use $\mathcal{T}^{00}$ to represent

In the HTL limit, the polarization functions $\Pi_T(p)$ and $\Pi_L(p)$ can we written in terms of  $\mathcal{T}^{00}$
\ba\la{tslo}
 \Pi_T(p)&=& \frac{m_D^2}{(d-1)n_p^2}\big[\mathcal{T}^{00}(p,-p)-1 +n_p^2 \big],  \nonumber \\  \Pi_L(p)&=& m_D^2\big[1-\mathcal{T}^{00}(p,-p)  \big]     \,.
\ea
Note that it is essential to take the angular average in $d=3-2\epsilon$ in (\ref{ts}), and then analytically continue to $d=3$ only after all poles in $\epsilon$ have been elimimated. The expression for $\mathcal{T}^{00}$ is
\ba\la{t00}
\mathcal{T}^{00}(p,-p)=\frac{\omega(\epsilon)}{2}\int^{1}_{-1}dc \, (1-c^2)^{-\epsilon} \frac{p_0}{p_0-|\textbf{p}|c} \, ,
\ea
where the weight function $\omega(\epsilon)$
\ba\la{t000}
 \omega(\epsilon)= \frac{\Gamma(2-2\epsilon)}{\Gamma^2(1-\epsilon)}2^{2\epsilon} =\frac{\Gamma( \frac{3}{2}  -\epsilon)}{\Gamma(\frac{3}{2})\Gamma(1-\epsilon)}    \,.
\ea
The integral in (\ref{t00}) must be defined so that it is analytic at $|p_0|=\infty$. It then has a branch cut running from $p_0=-|\textbf{p}|$ to $p_0=|\textbf{p}|$. If we take the limit $\epsilon\rightarrow 0$, it reduces to its $d=3$ form
\ba\la{t001}
\mathcal{T}^{00}(p,-p)= \frac{p_0}{2|\textbf{p}|} \log \frac{p_0+|\textbf{p}|}{p_0-|\textbf{p}|}  \,.
\ea

From the results above, we see that the definition of the gluon self energy in (\ref{sfgy}) and (\ref{sfg}) is the same as in QCD in Ref.~\cite{Andersen:2002ey} up to the definition of $m_D$.  Furthermore, as shown in Ref.~\cite{Czajka:2012gq}, this means that the HTL three-gluon vertex, four-gluon vertex, and ghost-gluon vertex are also the same as obtained in QCD after adjustment of $m_D$.

\subsection{Gluon propagator }

The Feynman rule for the gluon propagator is
\ba\la{prog}
i\delta^{ab}\Delta_{\mu\nu}(p)  \, ,
\ea
where the gluon propagator tensor $\Delta_{\mu\nu}$ depends on the choice of gauge fixing. In the limit $\xi\rightarrow \infty$, the its inverse reduces to
\ba\la{prog1}
\Delta^{-1}_\infty(p)^{\mu\nu}&=&-p^2g^{\mu\nu} +p^\mu p^\nu -\Pi^{\mu\nu}(p)    \nonumber \\ &=& -\frac{1}{\Delta_T(p)}T_p^{\mu\nu}  +\frac{1}{n_p^2\Delta_L(p)}L_p^{\mu\nu} \, ,
\ea
where $\Delta_T$ and $\Delta_L$ are the transverse and longitudinal propagators
\ba\la{prog2}
\Delta_T(p)&=&\frac{1}{p^2-\Pi_T(p)},   \nonumber \\  \Delta_L(p)&=&\frac{1}{-n_p^2p^2+\Pi_L(p)}   \,.
\ea
The inverse propagator for general $\xi$ is
\ba\la{prog3}
\Delta^{-1}(p)^{\mu\nu}=\Delta^{-1}_\infty(p)^{\mu\nu}-\frac{1}{\xi}p^\mu p^\nu \, ,
\ea
then by inverting the tensor $\Delta^{-1}(p)^{\mu\nu}$, we can get
\ba\la{prog4}
\Delta^{\mu\nu}(p)=-\Delta_T(p)T_p^{\mu\nu}+\Delta_L(p)n_p^{\mu}n_p^{\nu}-\xi \frac{p^\mu p^\nu}{(p^2)^2}\,.
\ea

In the course of the calculation it proved to be convenient to introduce the following propagators
\ba\la{prog5}
 \Delta_X(p)=\Delta_L(p)+\frac{1}{n_p^2}\Delta_T(p)\,.
\ea
Using (\ref{wis1}) and (\ref{prog2}), it can also be expressed as
\ba\la{prog6}
 \Delta_X(p)=\big[m_D^2-d \Pi_T(p)   \big]\Delta_L(p) \Delta_T(p) \, ,
\ea
which vanishes in the limit $m_D\rightarrow 0$. Using this form, gluon propagator tensor can be written as
\ba\la{prog7}
\Delta^{\mu\nu}(p)&=&\big[-\Delta_T(p)g^{\mu\nu}+\Delta_X(p)n^\mu n^\nu \big]-\frac{n\cdot p}{p^2}\Delta_X(p)(p^\mu n^\nu+n^\mu p^\nu) \nonumber \\ &+&\bigg[ \Delta_T(p)+\frac{(n\cdot p)^2}{p^2}\Delta_X(p)-\frac{\xi}{p^2}  \bigg]\frac{p^\mu p^\nu}{p^2}\,.
\ea

\subsection{Quark self-energy}\la{qsfn}

In $\mathcal{N}=4$ SYM theory, there are three diagrams that contribute to the quark self energy. In HTL limit, the quark self energy was computed in Ref.~\cite{Czajka:2012gq} for massless bosons and fermions
\be\la{sfqy}
\Sigma_{ab}^{ij}(p)=\frac{g^2}{2}N_c \delta_{ab}\delta^{ij} \int\frac{d^3 k}{(2\pi)^3} \frac{f(\textbf{k})}{|\textbf{k}|}\frac{{\displaystyle{\not} k}
}{p\cdot k}   \,.
\ee
This form is not complicated and we can divide $|\textbf{k}|$ directly for the last part in (\ref{sfqy}) giving
\be\la{sfqy1}
\Sigma_{ab}^{ij}(p)=\frac{g^2}{2}N_c \delta_{ab}\delta^{ij} \int\frac{k^2 dk}{2\pi^2} \frac{f(\textbf{k})}{|\textbf{k}|}\int\frac{d\Omega}{4\pi}\frac{{\displaystyle{\not} y}
}{p\cdot y}   \,,
\ee
after integration for momentum $|\textbf{k}|$, the HTL quark self energy can be written as
\be\la{sfq}
 \Sigma(p)=m_q^2{\displaystyle{\not} \mathcal{T}}(p),  \quad\quad m_q^2=\frac{1}{2}\lambda T^2   \, ,
\ee
where we have suppressed the trivial Kronecker deltas and
\be\la{sfq2}
 \mathcal{T}^\mu (p) \equiv \bigg\langle \frac{y^\mu}{p\cdot y} \bigg\rangle_{\hat{\textbf{y}}}\,,
\ee
and $m_q^2$ is the quark mass in super symmetry, satisfies $m_q^2=1/4 m_D^2$.

Similar to the gluon polarization tensor, the angular average in $ \mathcal{T}^\mu$ can be expressed as
\be\la{sfq3}
\mathcal{T}^\mu(p)=\frac{\omega(\epsilon)}{2}\int^{1}_{-1}dc(1-c^2)^{-\epsilon} \frac{y^\mu}{p_0-|\textbf{p}|c}      \,.
\ee
The integral in (\ref{sfq3}) must be defined, so that it is analytic at $|p_0|=\infty$. It then has a branch cut running from $p_0=-|\textbf{p}|$ to $p_0=|\textbf{p}|$. In three dimensions, it can be written as
\be\la{sfq4}
 \Sigma(p)=\frac{m_q^2}{2|\textbf{p}|}\gamma_0\log\frac{p_0+|\textbf{p}|}{p_0-|\textbf{p}|}+\frac{m_q^2}{|\textbf{p}|}\gamma\cdot \hat{\textbf{p}}\bigg(1-\frac{p_0}{2|\textbf{p}|}   \log\frac{p_0+|\textbf{p}|}{p_0-|\textbf{p}|}    \bigg)\,.
\ee

We can see that the definition of quark self energy in (\ref{sfq}) is the same as in QCD \cite{Andersen:2003zk} up to the definition of $m_q$ and taking into account that there are four Majorana fermions indexed by $i$.  In practice, this means that the quark propagator, quark-gluon three vertex and quark-gluon four vertex in HTLpt are the same as in QCD after the appropriate adjustment of the group structure constants.  We will take the results for these from Ref.~\cite{Andersen:2003zk} with the understanding that the finite-temperature quark mass should be understood to that of the SYM theory.

\subsection{Quark propagator}

The Feynman rule for the quark propagator is
\ba\la{proq}
i\delta^{ab}\delta^{ij}S(p)  \quad\quad  \textrm{with} \quad\quad   S(p)=\frac{1}{{\displaystyle{\not} p}-\Sigma(p)}  \, ,
\ea
where $i,j$ index the Majorana fermion being considered.  As a result, the inverse quark propagator can be written as
\ba\la{proq1}
S^{-1}(p)={\displaystyle{\not} p}-\Sigma(p)\equiv {\displaystyle{\not}\mathcal{A}}(p) \, ,
\ea
where $\mathcal{A}_\mu(p)=(\mathcal{A}_0(p),\mathcal{A}_s(p)\hat{\textbf{p}})$ with
\ba\la{proq2}
\mathcal{A}_0(p)&=&p_0-\frac{m_q^2}{p_0}\mathcal{T}_p \, ,     \nonumber \\
\mathcal{A}_s(p)&=&|\textbf{p}|+\frac{m_q^2}{|\textbf{p}|} \big[1-\mathcal{T}_p  \big]   .
\ea

\subsection{HTL quark counterterm}

The insertion of an HTL quark counterterm into a quark propagator is
\ba\la{proq3}
i\delta^{ab}\delta^{ij}\Sigma(p) \,.
\ea
where $\Sigma(p)$ is the HTL quark self energy given in (\ref{sfq}).

\subsection{Scalar self-energy} \la{ssfn}

 There are four diagrams that contribute to the scalar self energy in $\mathcal{N}=4$ SYM theory. In the HTL limit, the scalar self energy $\mathcal{P}_{ab}^{AB}$ was computed in Ref.~\cite{Czajka:2012gq} for massless bosons and fermions
\ba\la{sfsy}
\mathcal{P}_{ab}^{AB}(p)=g^2N_c\delta_{ab} \delta^{AB}  \int\frac{d^3 k}{(2\pi)^3} \frac{f(\textbf{k})}{|\textbf{k}|} \, .
\ea
After integration over the length of the three-momentum $|\textbf{k}|$, the HTL scalar self energy reduces to
\ba\la{sfsy1}
\mathcal{P}_{aa}^{AA}(p)=g^2N_cT^2=\lambda T^2=M_D^2 \, ,
\ea
where $M_D^2$ is the adjoint scalar mass, which has been given in Ref.~\cite{Czajka:2012gq,Kim:1999sg}. We can see that it satisfies $M_D^2=m_D^2/2=2m_q^2$.

Note that the scalar self energy is a constant, which means that it only affects the scalar propagator and not the scalar-gluon and scalar-quark vertices in $\mathcal{N}=4$ SYM theory. This is due to the fact that the HTL Lagrangian density $\mathcal{L}_{\textrm{HTL}}$ is a combination of the fields and their corresponding covariantized self energies and the HTL vertices are obtained by expanding the covariant derivatives $D_\mu$ appearing in the HTL effective Lagrangian in powers of the gauge field $A_\mu$. Since there are no covariant derivatives appearing in the scalar contribution to the HTL effective action \eqref{htl}, the scalar-gluon vertices will not receive corrections in HTLpt.\footnote{Ref.~\cite{Mrowczynski:2004kv} details the steps necessary to obtain the QCD HTLpt propagators and vertices from the QCD HTL effective action for both equilibrium and non-equilibrium systems.}$^,$\footnote{We are grateful for the authors of Ref.~\cite{Czajka:2012gq} for bringing this to our attention.}

\subsection{Scalar propagator}

The Feynman rule for the scalar propagator is
\ba\la{pros}
i \delta^{ab}\delta^{AB}\Delta_s(p) \,,
\ea
where
\ba\la{pros1}
\Delta_s(p)=\frac{1}{p^2-M_D^2} \,,
\ea
and its inverse is
\ba\la{pros2}
\Delta^{-1}_s(p)=p^2-M_D^2 \,.
\ea

\subsection{HTL scalar counterterm }

The insertion of an HTL scalar counterterm into a scalar propagator is
\ba\la{pros3}
-i \delta^{ab}\delta^{AB}\mathcal{P}_{aa}^{AA}(p)\,.
\ea
where $\mathcal{P}_{aa}^{AA}(p)$ is the HTL scalar self energy given in (\ref{sfsy1}).

\subsection{Quark-gluon vertex}
\label{subsec:qgv}

The quark-gluon vertex with incoming gluon momentum $p$, incoming quark momentum $r$, and outgoing quark momentum $q$, Lorentz index $\mu$, and color indices $a$, $b$, $c$ is
\ba\la{3qg}
\Gamma_{abc}^{\mu,ij}(p,q,r)&=& -g f_{abc}\delta^{ij}\big[\gamma^\mu + m_q^2\tilde{\mathcal{T}}^\mu(p,q,r)  \big] \nonumber \\   &=& -g f_{abc}\delta^{ij}\Gamma^{\mu}(p,q,r) \, .
\ea
Note that the sign on the second term differs from Ref.~\cite{Andersen:2003zk}.  This appears to be a typo in the original reference.
The rank-one tensor $\tilde{\mathcal{T}}^\mu$ in the HTL correction term is only defined for $p+r-q=0$
\ba\la{3qg2}
\tilde{\mathcal{T}}^\mu(p,q,r) =\bigg\langle y^\mu\bigg(\frac{{\displaystyle{\not} y}}{(y\cdot r)(y\cdot q)}  \bigg)     \bigg\rangle_{\hat{\textbf{y}}} \, ,
\ea
and is even under the permutation of $q$ and $r$. It satisfies the "Ward identity"
\ba\la{3qg3}
p_\mu \tilde{\mathcal{T}}^\mu(p,q,r) ={\displaystyle{\not} \mathcal{T}} (r)-{\displaystyle{\not} \mathcal{T}}(q) \, .
\ea
Note that the overall sign here differs from Ref.~\cite{Andersen:2003zk}.  This appears to be a typo in the original reference.
The quark-gluon vertex therefore satisfies the Ward identity
\ba\la{3qg4}
p_\mu \Gamma^\mu(p,q,r)=S^{-1}(q)-S^{-1}(r) \,.
\ea

\subsection{Quark-gluon four vertex}

The quark-gluon four vertex with outgoing gluon momentum $p$, $q$, incoming quark momentum $r$, and outgoing quark momentum $s$ is
\ba\la{4qg}
\Gamma_{abcd}^{\mu\nu,ij}(p,q,r,s)=-i g^2 \delta^{ij}m_q^2 \tilde{\mathcal{T}}^{\mu\nu}_{abcd}(p,q,r,s) \, ,
\ea
where we note that compared to QCD, the SYM theory has only quark indices in the adjoint representation. There is no tree-level term. The rank-two tensor $\tilde{\mathcal{T}}^{\mu\nu}$ is only defined for $p+q+s-r=0$
\ba\la{4qg1}
\tilde{\mathcal{T}}^{\mu\nu}_{abcd}(p,q,r,s)&=&f_{cde}f_{bae}\bigg\langle y^\mu y^\nu\frac{{\displaystyle{\not}y}}{(y\cdot r)(y\cdot s)[y\cdot(r-p)]}      \bigg\rangle \nonumber \\
&& + f_{bde}f_{cae}\bigg\langle y^\mu y^\nu\frac{{\displaystyle{\not}y}}{(y\cdot r)(y\cdot s)[y\cdot(s+p)]}      \bigg\rangle   ,
\ea
and satisfies
\ba\la{4qg2}
\delta^{ij}\delta^{ad}\delta^{bc}\Gamma_{abcd,ij}^{\mu\nu}(p,q,r,s) &=&-4i g^2N_c d_A \Gamma^{\mu\nu}(p,q,r,s)\,,
\ea
where
\be\la{4qg3}
\Gamma^{\mu\nu}(p,q,r,s)=m_q^2\bigg\langle y^\mu y^\nu  \bigg( \frac{1}{y\cdot r}+ \frac{1}{y\cdot s}   \bigg) \frac{{\displaystyle{\not}y}}{[y\cdot(r-p)][y\cdot(s+p)]}             \bigg\rangle .
\ee
This tensor is symmetric in $\mu$ and $\nu$, and satisfies the Ward identity
\be\la{4qg4}
p_\mu\Gamma^{\mu\nu}(p,q,r,s)= \Gamma^{\nu}(q,r-p,s)-\Gamma^{\nu}(q,r,s+p)\,.
\ee

\subsection{ Four-scalar vertex}

The four-scalar vertex does not depend on the momentum and is
\ba\la{4s}
\Gamma_{abcd}^{ABCD}(p,q,r,s)&=&-i g^2\bigg[ f_{abe}f_{cde}\bigg( \delta^{AC}\delta^{BD}-\delta^{AD}\delta^{BC}  \bigg) \nonumber \\ && +f_{ace}f_{bde}\bigg( \delta^{AB}\delta^{CD}-\delta^{AD}\delta^{BC}\bigg) \nonumber \\  && +f_{ade}f_{bce}\bigg( \delta^{AB}\delta^{CD}-\delta^{AC}\delta^{BD}    \bigg)   \bigg]\,.
\ea
This vertex satisfies
\ba\la{4s1}
\delta^{bd}\delta^{ac}\delta^{AC}\delta^{BD}\Gamma_{abcd}^{ABCD}(p,q,r,s)=(-i g^2)(60N_c d_A)\,,
\ea
where $\delta^{AA}=6$ for six scalars in this theory.

\subsection{Scalar-gluon vertex}

The scalar-gluon vertex with incoming gluon momentum $p$, incoming scalar momentum $r$, and outgoing scalar momentum $q$ is
\ba\la{sg}
 \Gamma_{abc}^{\mu,AB}(p,q,r)=g f_{abc}\delta^{AB}(r+q)^\mu\,.
\ea

\subsection{Scalar-gluon four vertex}

The scalar-gluon four vertex is independent on the direction of the momentum, and it can be expressed as
\ba\la{ssgg}
 \Gamma_{abcde}^{\mu\nu,AB}(p,q,r,s)=-2ig^2 g^{\mu\nu} \delta^{AB}f_{ade}f_{bce} \, ,
\ea
and satisfies
\ba\la{ssgg1}
 \delta^{ac}\delta^{bd}\delta^{AB}\Gamma_{abcde}^{\mu\nu,AB}(p,q,r,s)=(2ig^2) (6N_c d_A)g^{\mu\nu}\,.
\ea

\subsection{Quark-scalar vertex}

Since fermions have different interactions with the scalar ($X_{\texttt{p}}$) and pseudoscalar ($Y_{\texttt{q}}$) degrees of freedom, there are two kinds of vertex needed. One is quark-scalar vertex, with incoming scalar momentum $p$, outgoing quark momentum $q$ and incoming quark momentum $r$, and their corresponding colors $a,b,c$ respectively.  This vertex can be written as
\ba\la{qs1}
 \Gamma_{abc,ij}^{\texttt{p}}(p,q,r)=-i g f_{abc}\alpha_{ij}^{\texttt{p}}\,.
\ea
The other one is quark-pseudoscalar vertex
\ba\la{qs1a}
 \Gamma_{abc,ij}^{\texttt{q}}(p,q,r)= g f_{abc}\beta_{ij}^{\texttt{q}}\gamma_5\,.
\ea

\bibliographystyle{JHEP}
\bibliography{superhtl2}

\end{document}